\documentclass[useAMS,usenatbib]{mn2e}

\usepackage{times}
\usepackage{graphicx}
\usepackage{subfigure}

\newcommand{\gsim}{\mathrel{\hbox{\rlap{\lower.55ex \hbox {$\sim$}}
                   \kern-.3em \raise.4ex \hbox{$>$}}}}
\newcommand{\lsim}{\mathrel{\hbox{\rlap{\lower.55ex \hbox {$\sim$}}
                   \kern-.3em \raise.4ex \hbox{$<$}}}}

\title[The dependence of star formation on cloud structure]{The dependence of star formation on initial conditions and molecular cloud structure}

\author[M. R. Bate]
  {Matthew R. Bate\thanks{E-mail: mbate@astro.ex.ac.uk}\\
  $^1$School of Physics, University of Exeter, Stocker Road,
    Exeter EX4 4QL
}

\date{Accepted for publication in MNRAS}

\begin{document}
\maketitle

\begin{abstract}
We investigate the dependence of stellar properties on the initial kinematic structure of the gas in star-forming molecular clouds.  We compare the results from two large-scale hydrodynamical simulations of star cluster formation that resolve the fragmentation process down to the opacity limit, the first of which was reported by \citeauthor*{BatBonBro2003}.  The initial conditions of the two calculations are identical, but in the new simulation the power spectrum of the velocity field imposed on the cloud initially and allowed to decay is biased in favour of large-scale motions.  Whereas the calculation of \citeauthor{BatBonBro2003} began with a power spectrum $P(k)\propto k^{-4}$ to match the Larson scaling relations for the turbulent motions observed in molecular clouds, the new calculation begins with a power spectrum $P(k)\propto k^{-6}$.

Despite this change to the initial motions in the cloud and the resulting density structure of the molecular cloud, the resulting stellar properties resulting from the two calculations are indistinguishable.  This demonstrates that the results of such hydrodynamical calculations of star cluster formation are relatively insensitive to the initial conditions.  It is also consistent with the fact that the statistical properties of stars and brown dwarfs (e.g. the stellar initial mass function) are observed to be relatively invariant within our Galaxy and do not appear to depend on environment.
\end{abstract}


\section{Introduction}
\label{introduction}

The star formation process appears to be highly robust in producing stars and brown dwarfs with similar statistical properties regardless of variations in initial conditions and environment, at least within our Galaxy.  The primary product of the star formation process, the stellar initial mass function (IMF), does not appear to vary significantly \citep*{Scalo1998, Kroupa2002, Chabrier2003, ElmKleWil2008}.  Indeed, it is not uncommon for those who study star formation to question whether or not there is a `universal IMF', at least in the local Universe.  There have been some claims of variations, but these usually either appear less convincing on further investigation or can be plausibly explained by dynamical evolution.  For example, the claims of an under abundance of brown dwarfs in Taurus \citep{Bricenoetal1998, Luhman2000, Bricenoetal2002,Luhmanetal2003b} looked compelling at first sight, but the more one looks, the more brown dwarfs are found \citep{Luhman2004b, Guieuetal2006}.  The Arches cluster near the Galactic centre appears to have a top-heavy mass function \citep{Figeretal1999,Stolteetal2002}, but may also be explained by dynamical evolution \citep{Portegiesetal2002, Portegiesetal2007, Kimetal2006}.  Another case for a variation of the IMF is the young stars observed near the Galaxy's supermassive black hole, Sgr A$^*$ which appear to be deficient in low-mass stars \citep{NaySun2005}.  Other properties of the star formation process, for example, the frequencies and properties of binary systems also appear to be relatively consistent from region to region, but with a higher wide binary fraction in low-density star-forming regions \citep{Duchene1999,Reipurthetal2007}.  

On the theoretical side, the apparent invariance of the IMF and other stellar properties is a major obstacle to understanding the star formation process: if the results of the star formation process don't vary with some physical quantity, how can we determine which are the main processes involved in star formation?  It has long been thought that the characteristic stellar mass may originate from the typical Jeans mass in molecular clouds \citep[e.g.][]{Larson1978,Larson1992,Larson2005}.  This idea has been supported by hydrodynamical calculations of the fragmentation of clumpy and turbulent molecular clouds (\citealt*{KleBurBat1998}; \citealt{KleBur2000, KleBur2001, Klessen2001, Jappsenetal2005, BonClaBat2006}), including those that are able to resolve down to the opacity limit for fragmentation and, thus, capture the formation of all stars and brown dwarfs \citep*{BatBonBro2003,BatBon2005,Bate2005}.  \citet*{PadNorJon1997} link the characteristic stellar mass to the Jeans mass and the level of turbulence in the molecular cloud.  However, it is not immediately apparent how the dependence on the Jeans mass predicted by these models is consistent with the apparent invariance of the observed IMF.  Recently, \citet{ElmKleWil2008} proposed that the IMF does depend on the typical Jeans mass in a molecular cloud but, due to the thermodynamics of molecular gas, the typical Jeans mass only depends weakly on environment.  Another possibility \citep{Bate2009b} is that heating of the molecular cloud due to forming protostars alters the Jeans mass and self-regulates the star formation process, again weakening the dependence of the IMF on environment. 

In addition to the Jeans mass in molecular clouds, the question arises of whether the statistical properties of stellar systems depend on other physical variables.  \cite{Bate2005} found that changes in the opacity limit for fragmentation did not alter the IMF significantly, except at the low-mass cut-off.  However, some properties of the turbulence in molecular clouds have been found to affect the resulting IMF.  \citet{Klessen2001} found that small-scale driving of the turbulence produced a flattened IMF (i.e. the ratio of high-mass to low-mass stars increased).  \citet*{ClaBonKle2008} found that the IMF may also flatten if the kinetic energy in the molecular clouds is large enough to make them globally unbound which decreases the importance of competitive accretion.  Finally, \citet*{DelClaBat2004} and \citet*{GooWhiWar2006} investigated the dependence of star formation on the power spectrum of decaying turbulence in individual dense molecular cloud cores.  \citeauthor{DelClaBat2004} found that that initial conditions of $P(k)\propto k^{-3}$ rather than $P(k)\propto k^{-5}$ (i.e. more power in small-scale motions) gave more very-low-mass brown dwarfs (at the $2-\sigma$ level of significance) but left the rest of the IMF unchanged.  However, \citeauthor{GooWhiWar2006} found the opposite result (i.e. the fraction of low-mass objects increased when there was more power in large-scale motions).

In this paper, we report the results from a hydrodynamical calculation that resolves the collapse of large-scale molecular cloud down to the opacity limit for fragmentation that is initialised with decaying `turbulence' with a kinetic energy power spectrum biased towards large-scale motions of $P(k)\propto k^{-6}$.  The calculation is otherwise identical to that reported by \citet{BatBonBro2003}, which began with a kinetic energy power spectrum of $P(k)\propto k^{-4}$ to match the observed Larson scaling relations of the motions in molecular clouds \citep{Larson1981}.  Comparison of the results of these two calculations allows us to re-investigate the issue of the dependence of the star formation process on the kinetic power spectrum initially imposed on the gas.

The paper is structured as follows.  Section 2 briefly describes the numerical method and the initial conditions for the calculations.  The results are discussed in Section 3.  In Section 4, we discuss the implications of the results for the origin of the IMF.  Our conclusions are given in Section 5.

\section{Computational method}

The calculations presented here were performed using a three-dimensional, 
smoothed particle hydrodynamics (SPH) code.  The SPH code was 
based on a version originally developed by Benz \citep{Benz1990, Benzetal1990}.
The smoothing lengths of particles were variable in 
time and space, subject to the constraint that the number 
of neighbours for each particle must remain approximately 
constant at $N_{\rm neigh}=50$.  The SPH equations were 
integrated using a second-order Runge-Kutta-Fehlberg 
integrator with individual time steps for each particle 
\citep*{BatBonPri1995}.
Gravitational forces between particles and a particle's 
nearest neighbours were calculated using a binary tree.  
We used the standard form of artificial viscosity 
(Monaghan \& Gingold 1983; Monaghan 1992) with strength 
parameters $\alpha_{\rm_v}=1$ and $\beta_{\rm v}=2$.
Further details can be found in \citet{BatBonPri1995}.
The code was parallelised by M.\ Bate using OpenMP.

\begin{table*}
\begin{tabular}{lcccccccccc}\hline
Calculation & Initial & Initial  & Jeans & Mach & `Turbulent' Power 
& No. Stars & No. Brown  & Mass of Stars \&  & Mean & Median \\
 & Gas Mass  & Radius & Mass & No. & Spectrum Slope, $\beta$ 
 & Formed & Dwarfs  & Brown Dwarfs & Mass & Mass \\
 & M$_\odot$ & pc & M$_\odot$ & & $P(k)\propto k^{\beta}$ 
 & & & M$_\odot$ & M$_\odot$ & M$_\odot$\\ \hline
1 & 50.0 & 0.188 & 1 & 6.4 & $-4$ 
& $\geq$23 & $\leq$27 & 5.89 & 0.118 & 0.070 \\
4 & 50.0 & 0.188 & 1 & 6.4 & $-6$ 
& $\geq$20 & $\leq$22 & 6.29 & 0.150 & 0.073 \\
 \hline
\end{tabular}
\caption{\label{table1} The initial conditions for calculations 1 (BBB2003) and 4 (this paper) and the statistical properties of the stars and brown dwarfs formed.  The initial conditions for Calculation 4 were identical to those of Calculation 1.  The only difference was that the initial turbulent velocity fields had different power spectra, although they were both scaled so that the initial kinetic energy equalled the magnitude of the gravitational potential energy of the cloud.  Both calculations were run for 1.40 initial cloud free-fall times.  Brown dwarfs are defined as having final masses less than 0.075 M$_\odot$.  The numbers of stars (brown dwarfs) are lower (upper) limits because some of the brown dwarfs were still accreting when the calculations were stopped.}
\end{table*}

\subsection{Equation of state}
\label{eossec}

To model the thermal behaviour of the gas without performing radiative transfer,
we use a barotropic equation of state for the thermal pressure of the
gas $p = K \rho^{\eta}$, where $K$ is a measure of the entropy
of the gas.  The value of the effective polytropic exponent $\eta$, 
varies with density as
\begin{equation}\label{eta}
\eta = \cases{\begin{array}{rl}
1, & \rho \leq  10^{-13} {\rm g~cm}^{-3}, \cr
7/5, & \rho > 10^{-13} {\rm g~cm}^{-3}. \cr
\end{array}}
\end{equation}
We take the mean molecular weight of the gas to be $\mu = 2.46$.
The value of $K$ is defined such that when the gas is 
isothermal $K=c_{\rm s}^2$, with the sound speed
$c_{\rm s} = 1.84 \times 10^4$ cm s$^{-1}$ at 10 K,
and the pressure is continuous when the value of $\eta$ changes.
This equation of state has been chosen to match closely the 
relationship between temperature and density during the 
spherically-symmetric collapse of molecular 
cloud cores as calculated with frequency-dependent radiative 
transfer (see BBB2003 for further details).

\subsection{Sink particles}
\label{sinkparticles}

The heating of the molecular gas that begins at a density of $10^{-13}$ g~cm$^{-3}$ inhibits fragmentation at higher densities.  This is how we model the opacity limit for fragmentation.  The opacity limit for fragmentation results in the formation 
of distinct pressure-supported
fragments in the calculation.  As these fragments accrete, their
central density increases, and it becomes computationally impractical
to follow their internal evolution because of the short dynamical
time-scales involved.  Therefore, when the central density of 
a pressure-supported fragment exceeds 
$\rho_{\rm s} = 10^{-11}~{\rm g~cm}^{-3}$, 
we insert a sink particle into the calculation (Bate et al.\ 1995).

In all the calculations discussed in this paper, a sink particle is formed by 
replacing the SPH gas particles contained within $r_{\rm acc}=5$ AU 
of the densest gas particle in a pressure-supported fragment 
by a point mass with the same mass and momentum.  Any gas that 
later falls within this radius is accreted by the point mass 
if it is bound and its specific angular momentum is less than 
that required to form a circular orbit at radius $r_{\rm acc}$ 
from the sink particle.  Thus, gaseous discs around sink 
particles can only be resolved if they have radii $\gsim 10$ AU.
Sink particles interact with the gas only via gravity and accretion.

Since all sink particles are created from pressure-supported 
fragments, their initial masses are several M$_{\rm J}$, 
as given by the opacity limit for fragmentation
\citep{LowLyn1976,Rees1976,Silk1977a,Silk1977b,BoyWhi2005}.  
The lowest mass object produced by the four calculations 
discussed in this paper was $\approx 3$ Jupiter masses (M$_{\rm J}$).  
Subsequently, these fragments may accrete large amounts of material 
to become higher-mass brown dwarfs ($\lsim 75$ M$_{\rm J}$) or 
stars ($\gsim 75$ M$_{\rm J}$), but {\it all} the stars and brown
dwarfs begin as these low-mass pressure-supported fragments.

The gravitational acceleration between two sink particles is
Newtonian for $r\geq 4$ AU, but is softened within this radius
using spline softening \cite{Benz1990}.  The maximum acceleration 
occurs at a distance of $\approx 1$ AU; therefore, this is the
minimum separation that a binary can have even if, in reality,
the binary's orbit would have been hardened.  Sink particles are
not permitted to merge in this calculation.

The benefits and potential problems associated with introducing sink particles are discussed in more detail by \citet{BatBonBro2003}.

\subsection{Initial conditions}
\label{initialcond}

In this paper, we concentrate on the results from two calculations.  The initial conditions for the 
calculations (summarised in Table \ref{table1}) are identical, except for 
the initial `turbulent' velocity field that is imposed upon them.  
For each calculation, the initial conditions consist of a spherical
molecular cloud, uniform in density 
with a mass of 50 M$_\odot$.  The diameter of the clouds is 
$0.375$ pc (77400 AU).  At the temperature of 10 K, the mean thermal 
Jeans mass is 1 M$_\odot$ (i.e.\ each cloud contains 50 thermal 
Jeans masses).
The free-fall time of each cloud is $t_{\rm ff}=6.0\times 10^{12}$~s
or $1.90\times 10^5$ years.

In this paper, we refer to the two calculations as Calculations 1 and 4.  Results from Calculation 1 were presented in Bate et al.\ (2002a, 2002b, 2003).  Henceforth, we will refer to the latter of these papers as BBB2003.  We refer to the new calculation as Calculation 4 to differentiate it from Calculations 2 and 3 that were presented in \cite{BatBon2005} and \cite{Bate2005} (hereafter BB2005 and B2005), respectively.  These calculations were also of 50 M$_\odot$ molecular clouds, but investigated the dependence of the star formation on the initial mean thermal Jeans mass of the clouds and variations of the equation of state.

Although the clouds are uniform in density, we impose an initial 
supersonic `turbulent' velocity field on them in the same manner
as \cite{OstStoGam2001}.  We generate a
divergence-free random Gaussian velocity field with a power spectrum 
$P(k) \propto k^{\beta}$, where $k$ is the wavenumber.  In Calculations 1--3,
the slope of the power spectrum was set to $\beta=-4$.
In three dimensions, this results in a
velocity dispersion that varies with distance, $\lambda$, 
as $\sigma(\lambda) \propto \lambda^{1/2}$ in agreement with the 
observed Larson scaling relations for molecular clouds 
\citep{Larson1981}.
This power spectrum is slightly steeper than the Kolmogorov
spectrum, $P(k)\propto k^{-11/3}$.  Rather, it matches the 
amplitude scaling of Burgers supersonic turbulence associated
with an ensemble of shocks (but differs from Burgers turbulence
in that the initial phases are uncorrelated).
In the new calculation discussed in this paper, Calculation 4, we set $\beta=-6$.  This injects
more energy on large scales than the earlier calculations.  Our reason for
varying the power spectrum is two fold.  First, we wish to investigate the
dependence of the star formation process and the properties of the 
resulting stellar systems on the initial conditions of the 
velocity field.  Second, in some circumstances initial conditions
where most of the power is on large scales may be more realistic than 
those matching the Larson scaling relations.  
Specifically, if the molecular gas is swept up or perturbed
by an external source, such as the expansion of stellar winds, and HII region, or a supernova explosion, it is natural that the kinetic energy would be 
input on large scales and collapse may occur before a full turbulent cascade is established.

The velocity fields are generated on $64^3$ uniform grids and the velocities 
of the particles are interpolated from the grids with the diameter of the spherical 
molecular clouds set equal to the dimension of the grids (i.e., there are 64 grid 
cells across the diameter of each initial cloud).  Since each cloud initially contains 
50 Jeans masses, each Jeans mass measures roughly 16 grid cells across.  
Thus, the dynamic range of the initial turbulent velocity field is 64 while the 
dynamic range on scales larger than the initial Jeans length is $\approx 4$.  
Since the power spectrum slopes differ by two between the 
Calculations 1 and 4, the difference in power over a factor of 4 in 
scale (i.e. external to each Jeans mass) is at most $\approx 16$.  
While these numbers are not huge, if the star formation does depend 
sensitively on the initial turbulent power spectrum, one might expect 
to detect a difference in the properties of the stars that are formed
in the two calculations.

In each calculation, the velocity field is normalised so that the kinetic energy 
of the turbulence equals the magnitude of the gravitational potential energy of each cloud.
Thus, the initial root-mean-square (rms) Mach number of the turbulence 
is ${\cal M}=6.4$ for Calculations 1 and 4.  The velocity field is initialised
as described above and then allowed to decay as the calculation 
evolves; there is no `turbulent driving'.  In each case, the cloud is allowed to evolve freely 
into the vacuum surrounding it; there are no boundary conditions applied to the 
simulations.

\begin{table*}
\begin{tabular}{lccccccc}\hline
Core & Initial Gas  & Initial  & Final  & No. Stars & No. Brown  & Mass of Stars and  & Star Formation \\
 & Mass  & Size & Gas Mass & Formed & Dwarfs Formed & Brown Dwarfs & Efficiency  \\
 & M$_\odot$ & pc & M$_\odot$ &  & & M$_\odot$ & \% \\ \hline
1 & 1.78 (0.31) & $0.10\times 0.03\times 0.03$ & 5.66 (4.82) & $\geq$11 & $\leq$15 & 4.36 & 44 (47) \\
2 & 1.31 (0.27) & $0.05\times 0.03\times 0.03$ & 2.86 (2.19) & $\geq$9 & $\leq$7 & 1.93 & 40 (47)\\\hline
Cloud & 50.0 & $0.38\times 0.38\times 0.38$ & 43.7 & $\geq$20 & $\leq$22 & 6.29 & 13 \\ \hline
\end{tabular}
\caption{\label{table2} The properties of the two dense cores that form during Calculation 4 and those of the cloud as a whole.  The gas masses and sizes of the cores are calculated from gas with number density $n({\rm H}_2)>1\times 10^6$~cm$^{-3}$ and $n({\rm H}_2)>1\times 10^7$~cm$^{-3}$ (the latter values are given in parentheses).  The initial gas mass is calculated just before star formation begins in that core (i.e.\ different times for each core).  Brown dwarfs have final masses less than 0.075 M$_\odot$.  The star formation efficiency is taken to be the total mass of the stars and brown dwarfs that formed in a core divided by the sum of this mass and the mass in gas in that core at the end of the calculation.  As with Calculation 1, the star formation efficiency is high locally, but low globally.  The numbers of stars (brown dwarfs) are lower (upper) limits because fourteen of the brown dwarfs were still accreting when the calculation was stopped. }
\end{table*}

\subsection{Resolution}
\label{resolution}

The local Jeans mass must be resolved throughout the calculations to model fragmentation correctly 
\citep{BatBur1997, Trueloveetal1997, Whitworth1998, Bossetal2000, HubGooWhi2006}.  This requires $\gsim 1.5 N_{\rm neigh}$ SPH particles per Jeans mass; $N_{\rm neigh}$ is insufficient (BBB2003).  The minimum Jeans mass in Calculations 1 and 4 occurs at the maximum density during the isothermal phase of the collapse, $\rho_{\rm crit} = 10^{-13}$ g~cm$^{-3}$, and is $\approx 0.0011$ M$_\odot$ (1.1 M$_{\rm J}$).  Thus, we used $3.5 \times 10^6$ particles to model the 50-M$_\odot$ clouds.

Calculation 4 required approximately 100,000 CPU hours on the SGI Origin 3800 of the United Kingdom Astrophysical Fluids Facility (UKAFF).

\begin{figure*}
    \includegraphics[width=15.8cm]{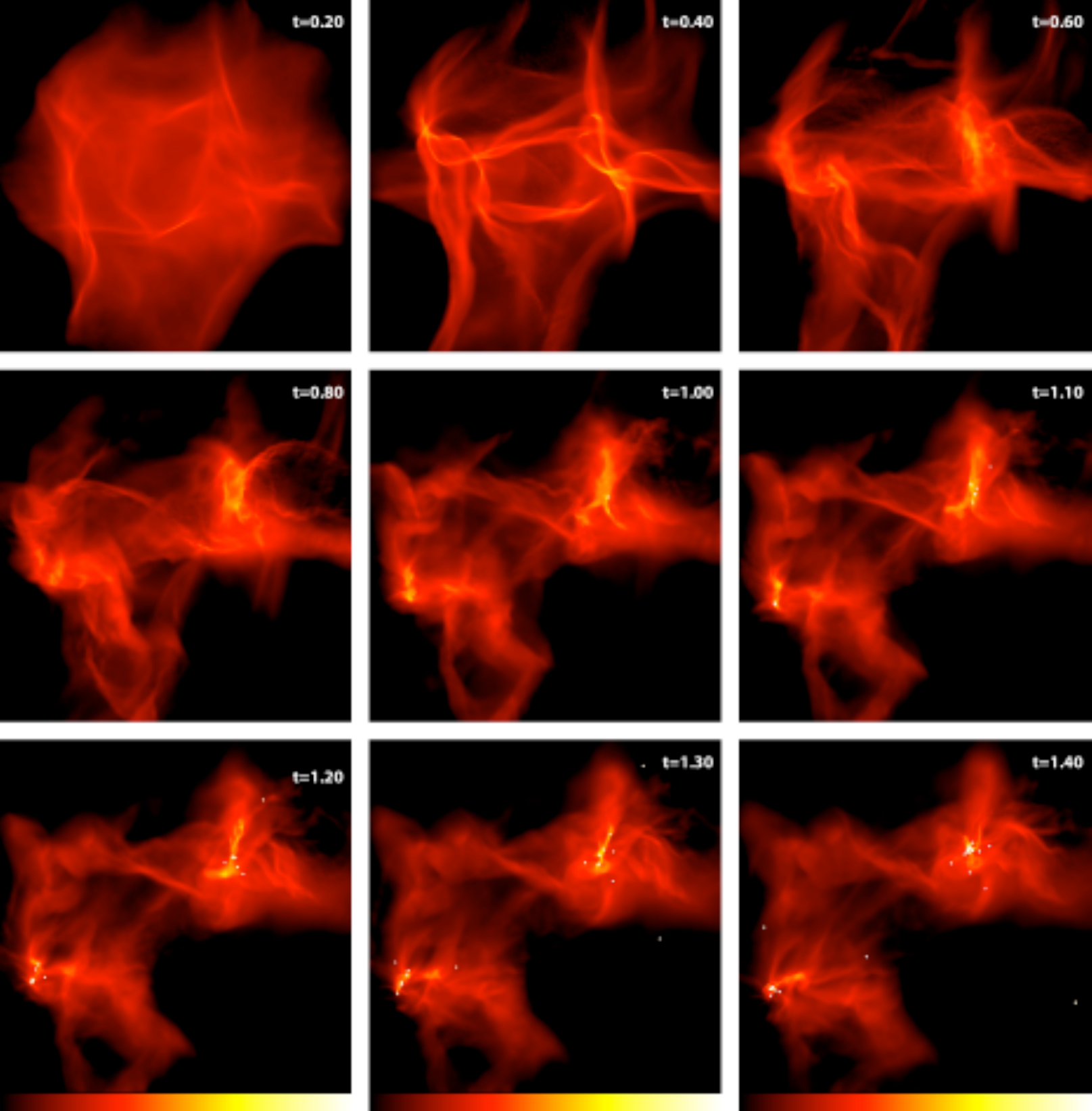}
\caption{\label{global} The global evolution of the cloud for comparison with Figure 2 of BBB2003 for Calculation 1.  By the end of the calculation, two dense cores have formed stars (bottom-left panel).  Many of the low-mass stars and brown dwarfs that formed in the dense cores have been ejected from the cores through dynamical interactions.  Each panel is 0.4 pc (82400 AU) across.  Time is given in units of the initial free-fall time of $1.90\times 10^5$ yr.  The panels show the logarithm of column density, $N$, through the cloud, with the scale covering $-1.7 < \log N < 1.5$ with $N$ measured in g cm$^{-2}$.} 
\end{figure*}

\section{Comparison of results}
\label{results}

The results of Calculations 1, 2 and 3 were published in BBB2003, BB2005, and B2005, respectively.  In these papers, the global evolution of the clouds, the star formation efficiencies and timescales, the forms of the stellar initial mass functions, the formation mechanisms of brown dwarfs and close binaries, the multiplicities and velocity dispersions of the objects, and the properties of their circumstellar discs were examined in detail.  In this paper, we present the results of Calculation 4 in an identical manner to the past calculations through the figures and tables, but in the text we concentrate on how the results {\it differ} from the other three calculations.  In particular, we concentrate on understanding the role of the initial velocity field in the progenitor molecular cloud in determining the statistical properties of the stars and brown dwarfs.

\begin{figure*}
    \includegraphics[width=15.8cm]{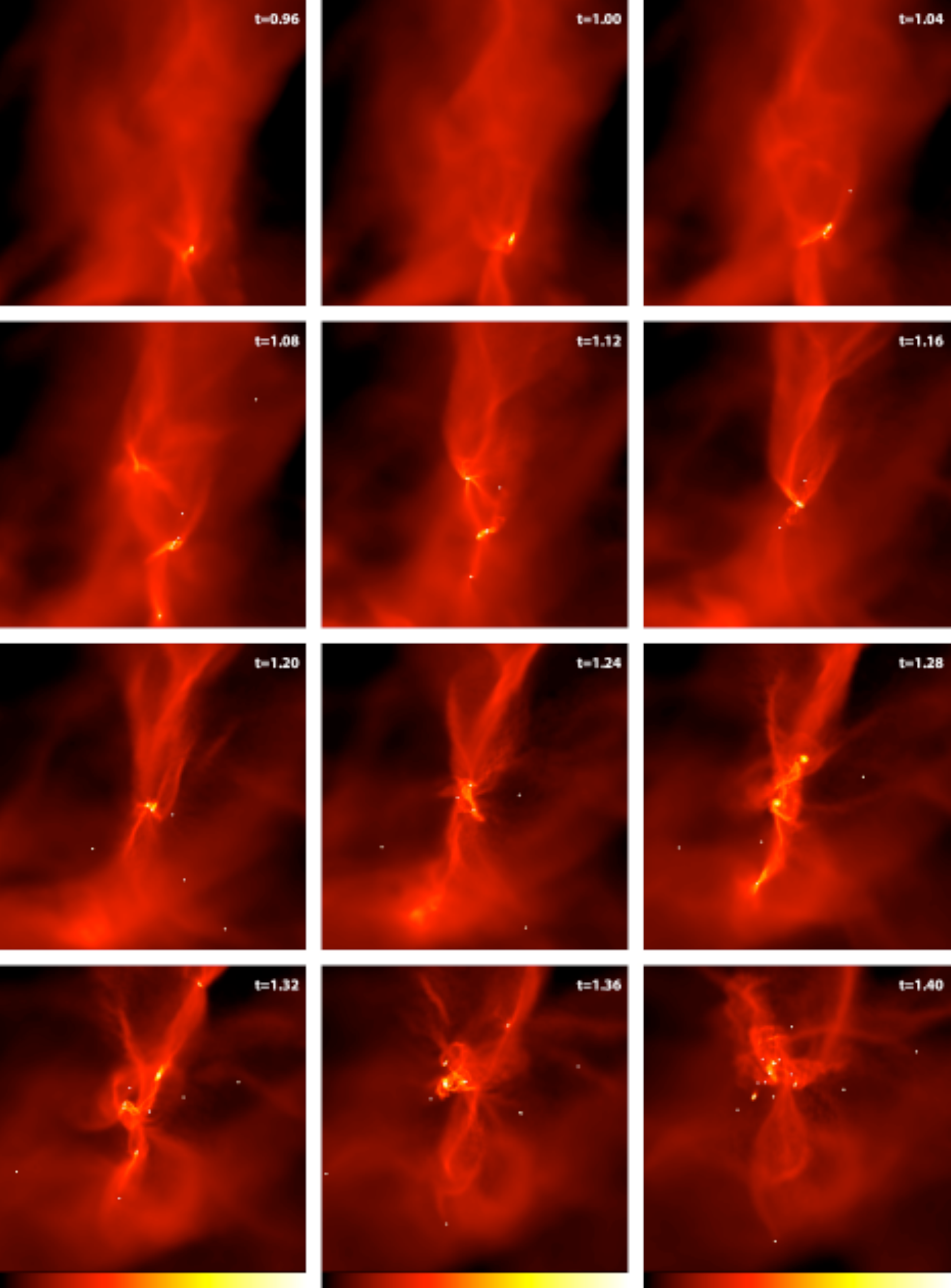}
\caption{\label{core1} The star formation in the first dense core.  The first object forms at $t=0.920~t_{\rm ff}$ in part of a long filament.  Its massive circumstellar disc later fragments into five objects $t=1.00-1.03~t_{\rm ff}$, the fourth of which is quickly ejected ($t=1.04, 1.08~t_{\rm ff}$ panels).  Two more objects form in the larger filament ($t=1.08, 1.12~t_{\rm ff}$ panels) and join the multiple system forming a small protostellar group ($t=1.16~t_{\rm ff}$ panel).  This group evolves by ejecting more objects ($t=1.20-1.36~t_{\rm ff}$) and is joined by another new object that forms in the filament (top of the $t=1.32~t_{\rm ff}$ panel).  Each panel is 0.0485 pc (10000 AU) across.  Time is given in units of the initial free-fall time of $1.90\times 10^5$ years.  The panels show the logarithm of column density, $N$, through the cloud, with the scale covering $-1.0 < \log N < 2.5$ with $N$ measured in g cm$^{-2}$.  } 
\end{figure*}

\begin{figure*}
    \includegraphics[width=15.8cm]{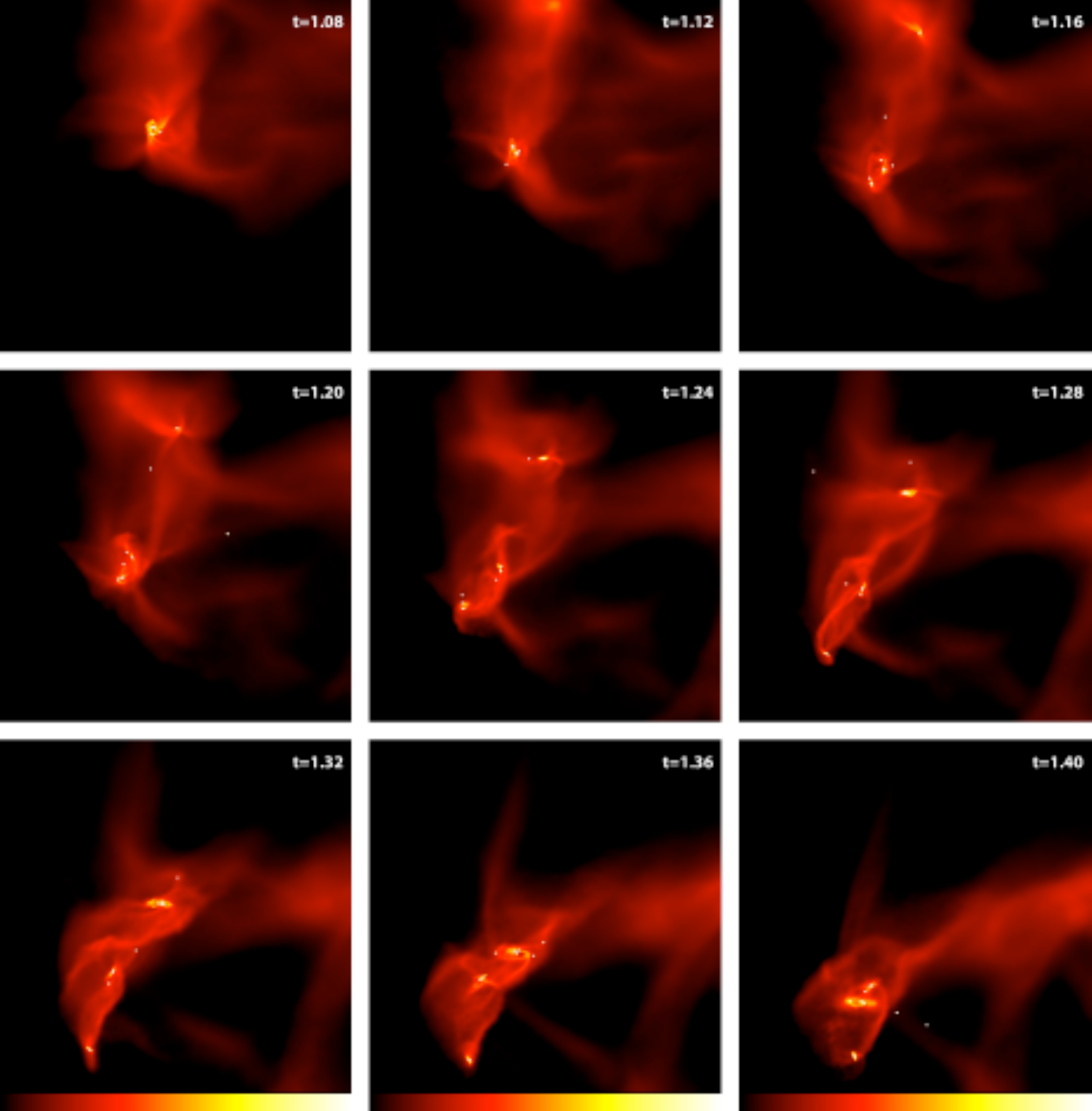}
\caption{\label{core2} The star formation in the second dense core.  The first objects form a binary at $t=1.049~t_{\rm ff}$ surrounded by a massive disc that subsequently fragments to produce another seven objects ($t=1.06-1.10~t_{\rm ff}$), two of which are ejected ($t=1.16, 1.20~t_{\rm ff}$ panels).  This multiple system is joined by another multiple system that forms $\approx 4600$ AU away (top of $t=1.16~t_{\rm ff}$ panel) as a single object with a massive disc that fragments into five objects ($t=1.27-1.33~t_{\rm ff}$).  Each panel is 0.0485 pc (10000 AU) across.  Time is given in units of the initial free-fall time of $1.90\times 10^5$ years.  The panels show the logarithm of column density, $N$, through the cloud, with the scale covering $-1.0 < \log N < 2.5$ with $N$ measured in g cm$^{-2}$. }
\end{figure*}

\subsection{Evolution of the cloud}
\label{evolution}

As with Calculations 1--3, due to the initial velocity dispersion of the gas the cloud quickly develops shocks, simultaneously losing kinetic energy and developing overdensities in regions with converging gas flows.    When gravity begins to dominate in an overdense region, gravitational collapse occurs and star formation begins.  However, because most of the kinetic energy is input on large scales in Calculation 4, its appearance for the first $0.5~t_{\rm ff}$ ($9.5 \times 10^4$~yr) is quite different from the other calculations (compare Figure \ref{global} with Figure 2 in BBB2003, and Figure 1 in each of BB2005, and B2005).  Whereas Calculations 1--3 produced abundant small scale structure initially, Calculation 4 produces long ($\approx 0.1-0.2$ pc) filaments by $t=0.5~t_{\rm ff}$.  These long filaments subsequently merge and break up to produce two main dense cores that undergo gravitational collapse.  However, throughout the remainder of the calculation, these one of these two cores in particular retains an underlying structure dominated by a single filament.

In Calculation 1, identical to Calculation 4 except for the initial velocity field, three dense star-forming cores were formed, one with a final mass of $\approx 8$~M$_\odot$ and two with final masses of $\approx 1.5$~M$_\odot$.  Calculation 4 produces two dense cores, the first with a final mass of $\approx 10$~M$_\odot$ and the second with a final mass of $\approx 5$~M$_\odot$ combining their gas and stellar masses (see Table \ref{table2}).  The two dense cores are separated by $\approx 0.3$ pc.  Whereas star formation began in Calculation 1 at $t=1.037~t_{\rm ff}$ ($1.97\times 10^5$~yrs), Calculation 4 begins producing stars slightly earlier at $t=0.920~t_{\rm ff}$ ($1.75\times 10^5$~yrs) in the first dense core and $t=1.048~t_{\rm ff}$ ($1.99\times 10^5$~yrs) in the second dense core.  By the end of the calculation, the first dense core has produced 26 objects and the second dense core 16 objects (Table \ref{table2}).

As with the previous calculations, Calculation 4 was stopped at $t=1.40~t_{\rm ff}$ to allow direct comparison of the results.  Star formation would continue in the cloud if the calculation was followed further.  Calculation 4 produced 20 stars and 18 brown dwarfs.  Four additional objects had substellar masses when the calculation was stopped but were still accreting.  It is impossible to tell whether or not they would become stars without continuing the calculation further.

\subsection{The star formation process in the dense cores}
\label{process}

Snapshots of the process of star formation in Calculation 4 are shown in Figure \ref{core1} for Core 1 and in Figure \ref{core2} for Core 2.  As with the earlier calculations, a true appreciation of how dynamic and chaotic the star-formation process is can only be obtained by studying an animation of the simulation.  The reader is encouraged to download an animation comparing Calculations 1 and 4 from http://www.astro.ex.ac.uk/people/mbate/Research/Cluster.

The star formation in the dense cores proceeds via gravitational collapse of filamentary structures (\citealt{Bastien1983, Bastienetal1991, InuMiy1992}; BBB2003) to form a combination of single objects and multiple systems (Figures \ref{core1} and \ref{core2}).  In Calculation 1 (BBB2003), many of the multiple systems resulted from the fragmentation of massive circumstellar discs \citep[e.g.,][]{Bonnell1994, BonBat1994a, BonBat1994b, Whitworthetal1995, BurBatBod1997, Hennebelleetal2004, BatBonBro2002a}.  This is also the case in Calculation 4.

The main difference between the star formation in the two dense cores of Calculation 4 and the main dense core of Calculation 1, which are all similar in mass, is that in Calculation 4 the star formation proceeds along a single filament in each of the dense cores (particularly in the first dense core; Figure \ref{core1}). In Calculation 1, the pattern of star formation is more `two dimensional' with multiple intersecting filaments, `sheets', and large discs. Thus, memory of the initial conditions is retained during the evolution and the large-scale perturbations applied in the initial conditions have an effect on the star formation process throughout Calculation 4, at least in terms of the distribution of star formation locations.

However, in both Calculations 1 and 4, protostars fall together along the filaments into the gravitational potential well of the core to form small stellar clusters (e.g., Figure \ref{core1}, $t=1.12-1.16~t_{\rm ff}$ and $t=1.28-1.40~t_{\rm ff}$).  Competitive accretion \citep{Bonnelletal1997}, chaotic dynamical interactions between objects, and ejections then play a similar role in both calculations.

\subsection{Star formation timescale and efficiency}
\label{efficiency}

The timescale on which star formation occurs is the dynamical one in all four calculations, consistent both with observational and other theoretical arguments \citep{Pringle1989, Elmegreen2000b, HarBalBer2001}, although we note that if strong magnetic fields were present they might slow the star formation rate by factors of a few \citep{PriBat2008,PriBat2009}.  We note that Calculations 1 and 4 convert similar amounts of gas into stars in the same amount of time (5.89 and 6.29 M$_\odot$, respectively).  Thus, the different power spectrum of the initial velocity field seems to have little effect on the star formation rate.  Both calculations have, of course, received the same amount of kinetic energy due to the normalisation of the velocity field.  \cite{ClaBon2004} have shown that if such star formation calculations are initialised with more kinetic energy, the star formation efficiency decreases.

By the end of all four calculations, most of the gas is in low-density regions where no star formation occurs.  Thus, the overall star formation efficiencies are low ($\sim 10$\%) for all calculations when they are stopped.  Although none of the calculation have been followed until star formation ceases, in all calculations a large fraction of the gas has drifted off to large distances by the end of the calculations due to the initial velocity dispersion and pressure gradients and is not gravitationally unstable.  Thus, the global star formation efficiencies are unlikely to exceed a few tens of percent.

\begin{figure}
    \includegraphics[width=8.7cm]{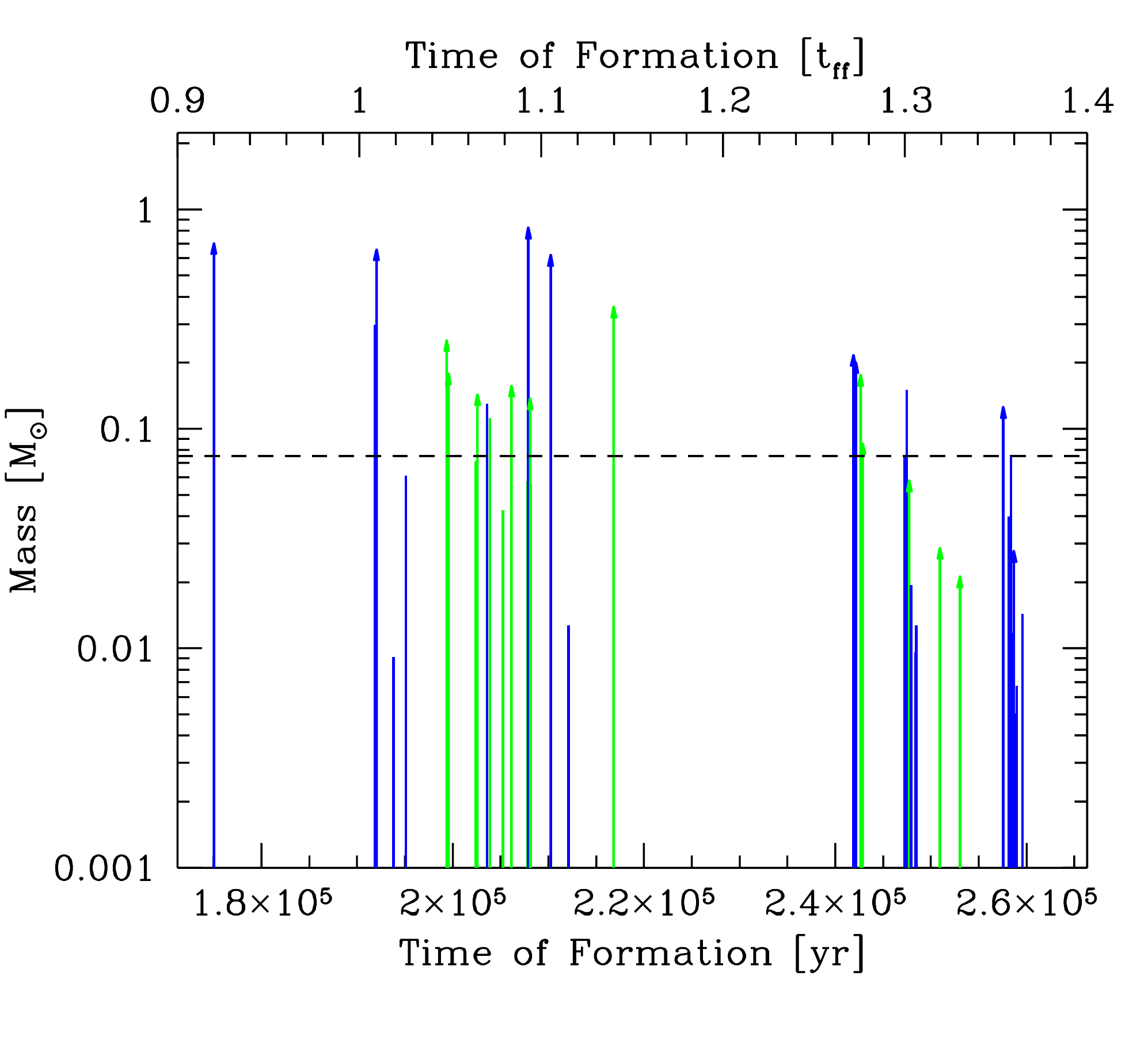}
\caption{\label{sfrate} Time of formation and mass of each star and brown dwarf 
at the end of the calculation.  The colour of each line identifies the dense core 
in which the object formed: first (blue), or second (green) core.
Objects that are still accreting significantly at the end of the calculation are
represented with arrows.  The horizontal dashed line marks the star/brown dwarf 
boundary.  Time is measured from the beginning of the calculation in terms of
the free-fall time of the initial cloud (top) or years (bottom).  This figure may be 
compared with the equivalent figures for Calculations 1--3 contained in 
BBB2003, BB2005, and B2005, respectively.} 
\end{figure}

However, in all calculations, the local star-formation efficiency is high within each of the dense cores (see Table \ref{table2} for Calculation 4).  This high star-formation efficiency is responsible for the bursts of star formation seen in all four calculations.   In Calculation 4 (see Figure \ref{sfrate}), following the formation of the first object at $t=0.92 t_{\rm ff}$, there is a burst of star formation in the first dense core from $t=1.01-1.11 t_{\rm ff}$, followed by a pause, and a second burst during $t=1.27-1.37 t_{\rm ff}$).  Similarly, the second dense core undergoes its first burst from $t=1.04-1.15 t_{\rm ff}$ and its second burst from $t=1.27-1.33 t_{\rm ff}$.  Gas is rapidly converted into stars in the dense cores and depleted to such an extent that star formation pauses.  Fresh gas must fall into the gravitational potential wells of the small clusters before new bursts of star formation can ensue.  Recently, \cite{Bate2009b} performed calculations that included radiative feedback from the protostars.  He found that this almost entirely shut off the second burst of star formation found in the purely hydrodynamical calculations.  Instead, the gas falling into the dense cores was primarily accreted by the protostars formed in the initial burst.  Although including radiative feedback has a significant effect on the outcome of the calculations, our intention here is simply to investigate whether changing the initial turbulent power spectrum has an effect on the outcome of the calculations.  It does not appear to affect the star formation rate significantly.

\subsection{Stellar velocity dispersion}
\label{velocitydispersion}

Dynamical interactions between cluster members eject stars and brown dwarfs in all four calculations.  In both Calculations 1 and 2, BBB2003 and BB2005 found that there was no significant dependence of the final velocity dispersion of the stars and brown dwarfs on either stellar mass or binarity.  In Calculation 3, binaries were found to have a somewhat smaller velocity dispersion than single objects.  While the lack of dependence on mass was also reported from past $N$-body simulations of the breakup of small-$N$ clusters with $N>3$ \cite{SteDur1998} and SPH calculations of $N=5$ clusters embedded in gas \citep{DelClaBat2003}, these calculations found that binaries should have a smaller velocity dispersion than single objects due to the recoil velocities of binaries being lower, keeping them within the stellar groups.  On the other hand, \cite{Delgadoetal2004} performed simulations of star formation in small turbulent clouds and found that the velocity dispersions of singles and binaries were indistinguishable, but that higher-order multiples had significantly lower velocity dispersions.   Recently, \cite{Bate2009a} performed a calculation similar to Calculation 1, but of a cloud an order of magnitude more massive (500~M$_\odot$) that produced more than 1250 stars and brown dwarfs.  With the accurate statistics provided by this large number of objects, \citeauthor{Bate2009a} was able to confirm that binaries do indeed have a lower velocity dispersion than single objects -- only 2/3 that of single objects.  He also found that very-low-mass objects (masses $<0.1$~M$_\odot$) had a slightly lower (80\%) velocity dispersion compared with higher-mass objects.

\begin{figure}\vspace{-1cm}
    \includegraphics[width=8.8cm]{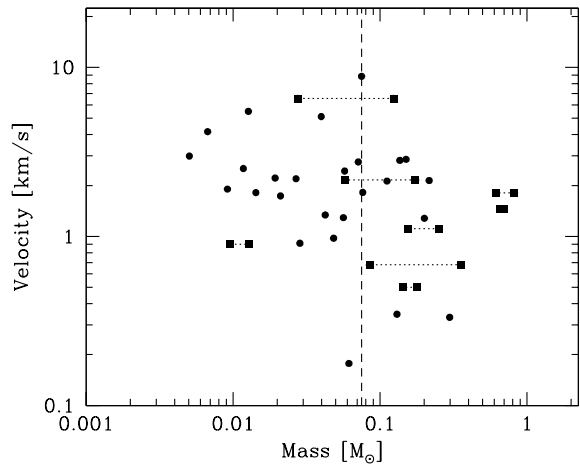}\vspace{-3cm}
\caption{\label{veldisp}  The velocities of each star and brown dwarf relative to
the centre-of-mass velocity of the stellar system.  For binaries with semi-major 
axes $< 100$ AU, the centre-of-mass velocity of the binary is given, and the
two stars are connected by dotted lines and plotted as squares rather than circles.  
The root mean square velocity dispersion 
for the association (counting each binary once) is 2.9 km/s (3-D) or 1.7 km/s (1-D).  As in Calculations 1 and 2, there is no significant dependence of the 
velocity dispersion on mass or binarity.
The vertical dashed line marks the star/brown dwarf boundary.} 
\end{figure}

The velocities of the stars and brown dwarfs relative to the centre of mass of all the objects are given in Figure \ref{veldisp} for Calculation 4.  The rms velocity dispersion is 2.9 km~s$^{-1}$ in three dimensions or 1.7 km~s$^{-1}$ in one dimension (using the centre-of-mass velocity for binaries with semi-major axes $<100$~AU).  This is intermediate between the velocity dispersions of Calculations 1--3, which had three-dimensional velocity dispersions of 2.1, 4.3, and 3.7 km~s$^{-1}$, respectively.  

The three-dimensional velocity dispersions of brown dwarfs, stars, and binaries (semi-major axes $< 100$ AU) are 3.4, 1.9, and 2.6 km~s$^{-1}$, respectively.  The difference between the velocity dispersions of the stars and brown dwarfs is not significant due to the small number of objects and the fact that the high brown dwarf velocity dispersion is partly based on a single object that lies right at the brown dwarf/star boundary ($0.075$~M$_\odot$) that was ejected with a velocity of 9 km~s$^{-1}$ (removing this object from the sample gives a brown dwarf velocity dispersion of 2.7 km~s$^{-1}$).  There is also no significant difference between velocity dispersion of the singles and binaries.  These results are consistent with the results obtained from the previous similar calculations.  They are also consistent with the results of \cite{Bate2009a} in that the differences found by \citeauthor{Bate2009a} with his more accurate statistics were small and would be impossible to detect given the small number of objects produced by Calculations 1 and 4. 

Observationally, there is no evidence for brown dwarfs having a significantly different velocity dispersion than stars.  \cite{ReiCla2001} had suggested that a greater velocity dispersion for brown dwarfs than stars may be a possible signature that brown dwarfs form as ejected stellar embryos.  In fact, in agreement with the calculations of \cite{Bate2009a}, studies of the radial velocities of stars and brown dwarfs in the Chamaeleon I dark cloud find that brown dwarfs have a marginally lower velocity dispersion than the T Tauri stars \citep{JoeGue2001,Joergens2006}. 

\begin{figure*}\vspace{-0.0cm}
    \includegraphics[width=8.8cm]{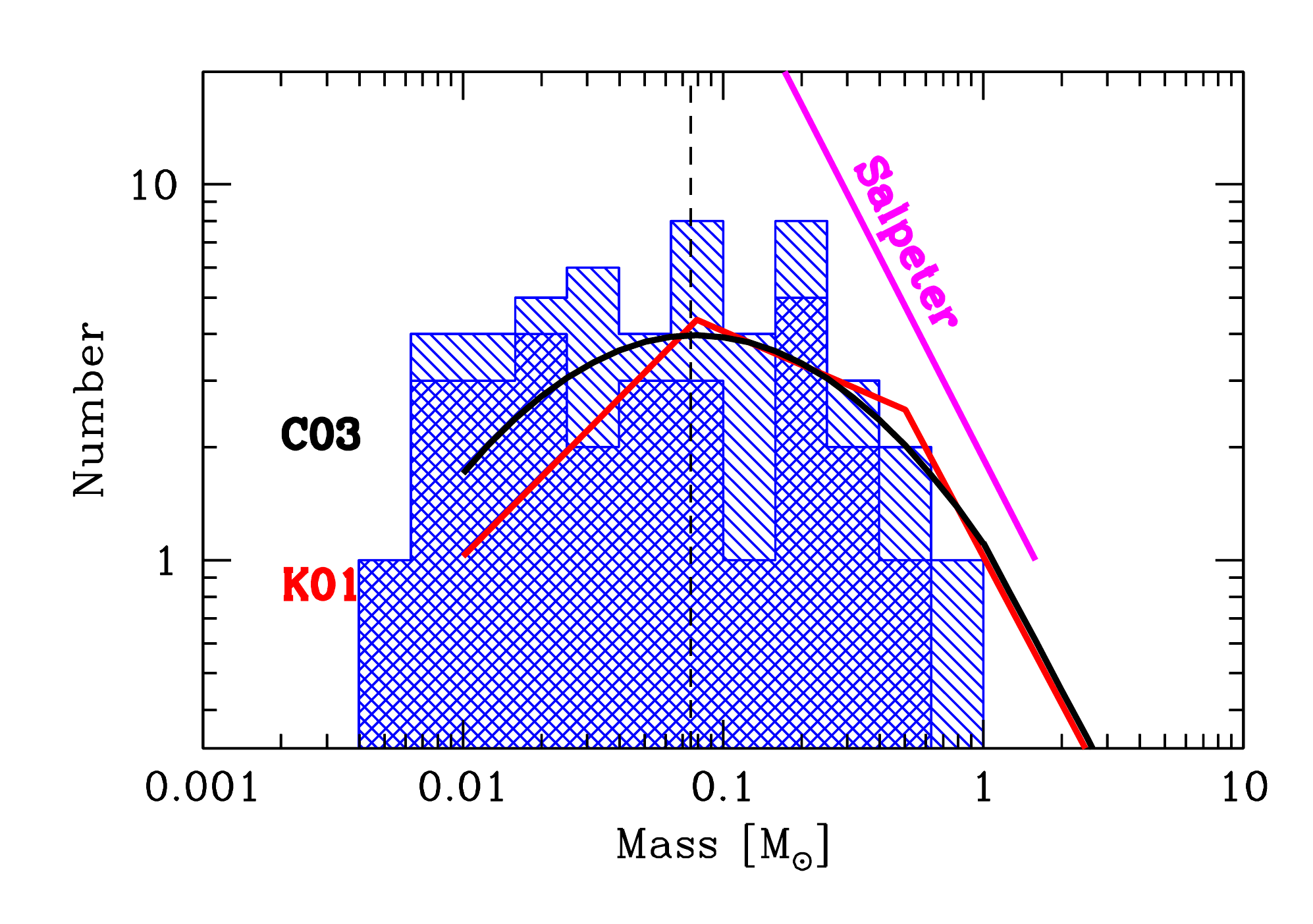}
    \includegraphics[width=8.8cm]{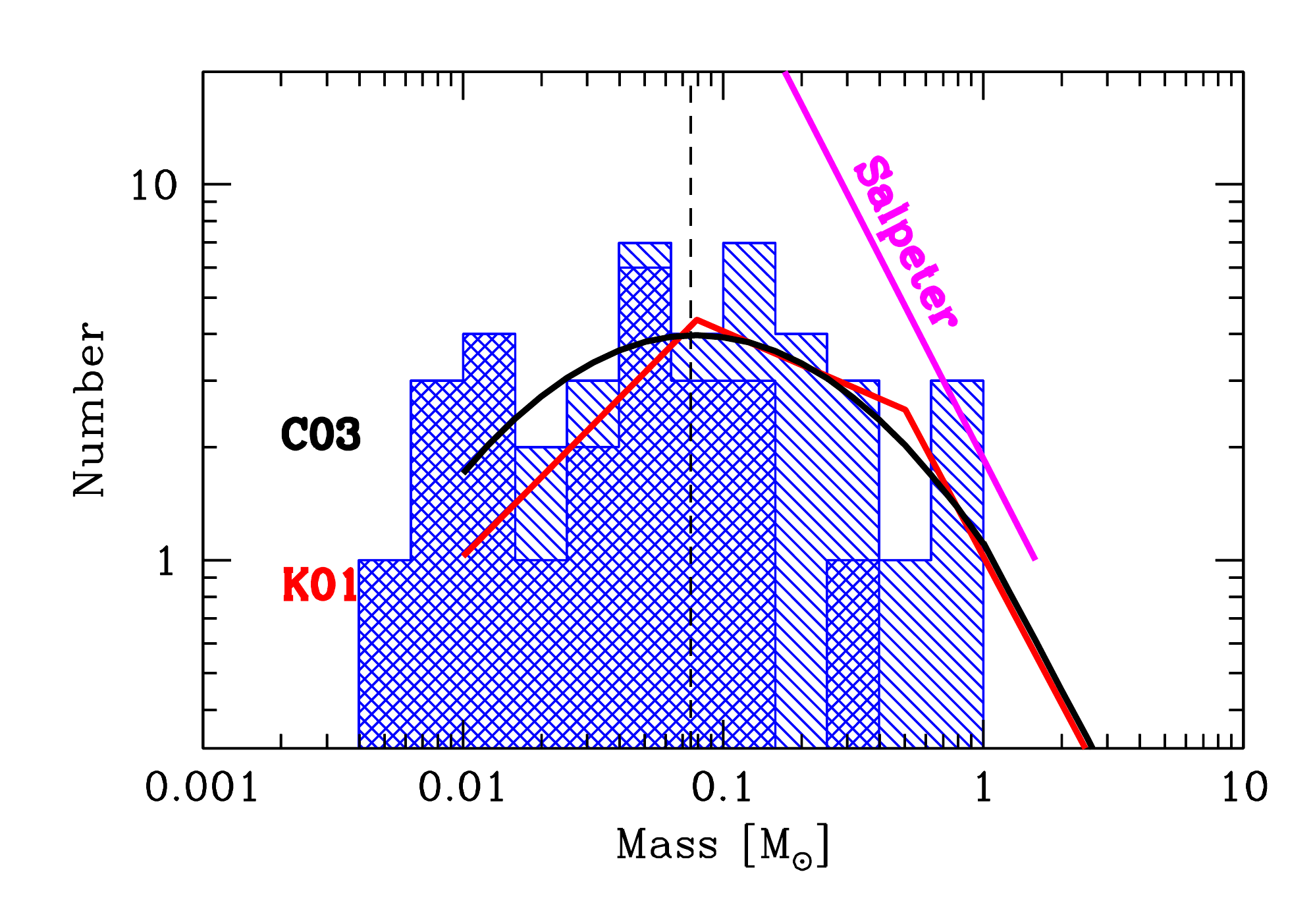}
\caption{\label{imf} The initial mass functions produced by Calculations 1 and 4.  Calculation 4 (right panel) is identical to Calculation 1 (left panel), except that the velocity field imposed on the gas initially had a steeper power spectrum that preferentially injected kinetic energy on larger scales.  In each case, the single shaded region shows all of the objects, the double shaded region shows only those objects that had finished accreting by the end of the calculations.  The mass resolution of the simulations is 0.0011 M$_\odot$ (i.e.\ 1.1 M$_{\rm J}$), but no objects have masses lower than 4.9~M$_{\rm J}$ in Calculation 1 and 5.0~M$_{\rm J}$ in Calculation 4 due to the opacity limit for fragmentation.  We also plot fits to the observed IMF from Kroupa (2001) (solid broken line), and Chabrier (2003) (solid curve).  The Salpeter (1955) slope (solid straight line) is equal to that of Kroupa (2001) for $M>0.5$ M$_\odot$.  The vertical dashed line marks the star/brown dwarf boundary.} 
\end{figure*}

\subsection{Initial mass function}
\label{imfsec}

A summary of the mass distributions of the stars and brown dwarfs formed in Calculations 1 and 4 is given in Table \ref{table1}.  From Calculations 1 and 2, BB2005 found that decreasing the mean thermal Jeans mass of the progenitor cloud by a factor of three resulted in a corresponding decrease in the median (characteristic) mass by a factor of almost exactly a factor of three.  Thus, they concluded that the characteristic stellar mass may be set by the mean thermal Jeans mass in molecular clouds.  B2005 investigated the dependence of the characteristic mass on the opacity limit for fragmentation.  He found that changing the opacity limit for fragmentation only altered the value of the minimum mass cut-off and did not alter the rest of the IMF significantly.  Here we investigate the dependence of the IMF on the power spectrum of the initial velocity field used in the calculations.

\begin{figure}\vspace{-1cm}
    \includegraphics[width=8.8cm]{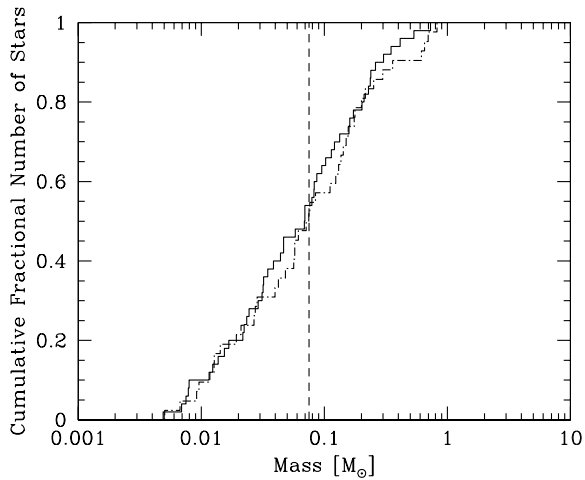}\vspace{-3.5cm}
\caption{\label{cumimf} The cumulative initial mass functions produced by Calculations 1 (solid line) and 4 (dotted line).  A Kolmogorov-Smirnov test on the two distributions gives a result of 95\%, showing that the two distributions are more similar than would be expected from random sampling of the same underlying IMF (i.e. statistically, they are indistinguishable).  The vertical dashed line marks the star/brown dwarf boundary.} 
\end{figure}

The only difference between Calculations 1 and 4 is the slope of the power spectrum of the initial velocity field.  The IMFs from the two calculations are given in Figure \ref{imf}.  Both calculations form roughly equal numbers of stars and brown dwarfs, indicating that changing the velocity power spectrum has little affect on the IMF.  In Calculation 1, 50 objects were formed with a mean object mass of 0.118 M$_\odot$ and a median mass of 0.070 M$_\odot$.  In Calculation 4, 42 objects were formed in the same time with a mean mass of 0.150 M$_\odot$ and a median mass of 0.073 M$_\odot$ (Table \ref{table1}).   Again, these numbers indicate that there is little difference between the two IMFs, however, to test this hypothesis we perform a Kolmogorov-Smirnov (K-S) test on the two cumulative IMFs (Figure \ref{cumimf}).  The K-S test on the two distributions shows that they are consistent with having been drawn from the same underlying IMF with the test giving a 95\% probability (i.e., statistically, they are indistinguishable).  By contrast, BB2005 found that the IMFs from Calculations 1 and 2 had only a 1.9\% probability of being drawn from the same underlying IMF.  Thus, we conclude that the IMF is insensitive to variations in the initial power spectrum of the decaying `turbulence' used in these calculations.

In Figure \ref{imf}, we compare the IMFs from Calculations 1 and 4 with parameterizations of the observed Galactic IMF by \cite{MilSca1979}, \cite{Kroupa2001}, and \cite{Chabrier2003}.  Given the small number statistics, both Calculations 1 and 4 are in agreement with Chabrier's single star IMF (black solid curve).

BB2005 investigated in detail the origin of the IMF from Calculations 1 and 2.  Since all objects in the calculations begin as opacity limited fragments at the minimum mass (a few Jupiter masses) and then accrete to their final masses, low-mass objects could originate from objects with low accretion rates or from objects with a typical accretion rate whose accretion is terminated shortly after they form (e.g., by ejection in a dynamical interaction with other objects).  

\begin{figure}
    \includegraphics[width=8.8cm]{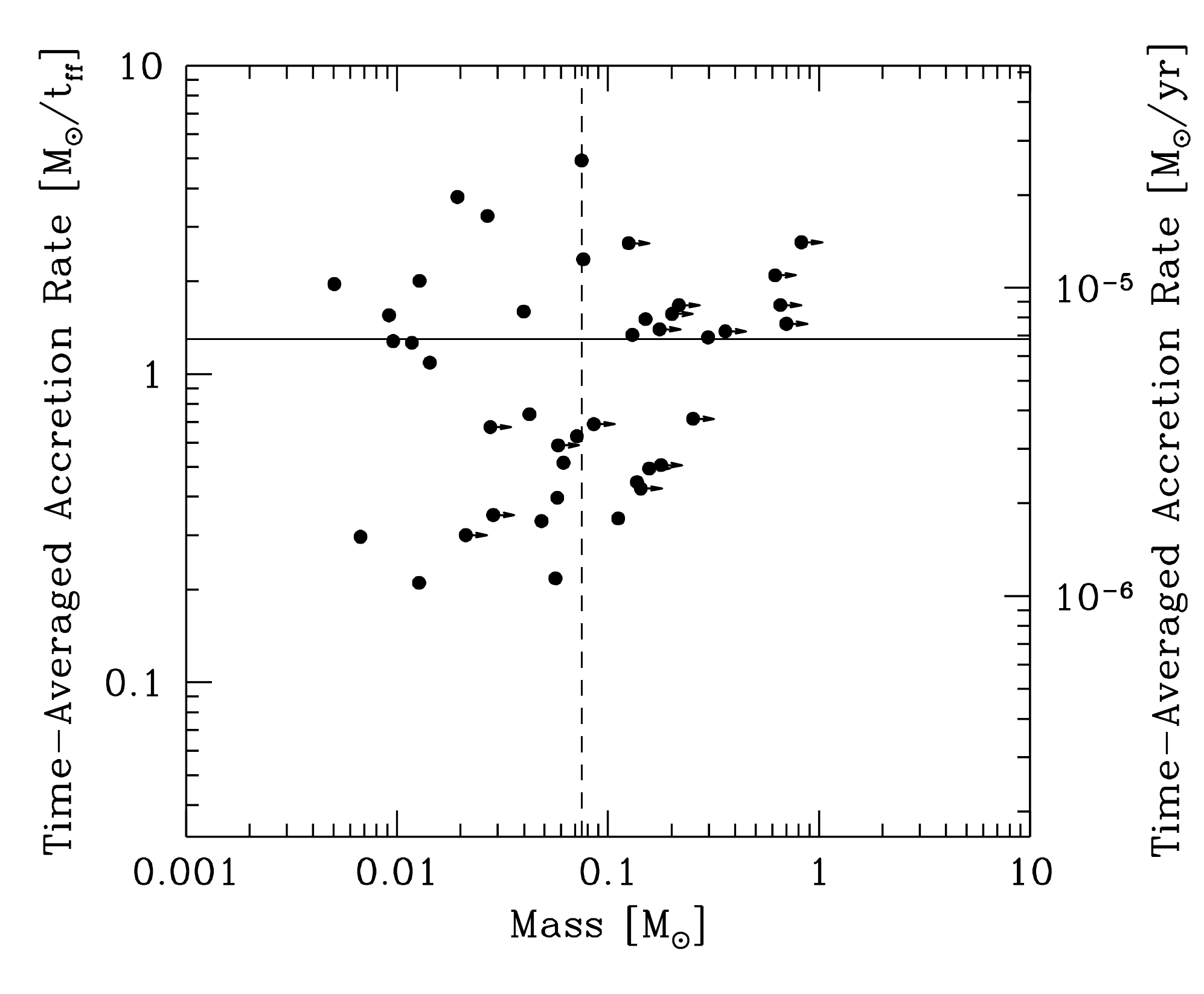}
\caption{\label{accrates} The time-averaged accretion rates of the objects formed in the calculation versus their final masses.  The accretion rates are calculated as the final mass of an object divided by the time between their formation and the termination of their accretion
or the end of the calculation.  The horizontal solid line gives the arithmetic
mean of the accretion rates: $6.6\times 10^{-6}$ M$_\odot$/yr.  The accretion rates are given in M$_\odot/t_{\rm ff}$ on the left-hand axes
and M$_\odot$/yr on the right-hand axes. The vertical dashed line marks the star/brown dwarf boundary.} 


\end{figure}

Following BB2005, in Figure \ref{accrates}, we plot the time-averaged accretion rates of all the objects in Calculation 4 as a function of their final masses.  A time-averaged accretion rate is defined as the mass of an object at the end of the calculation divided by the time over which it accreted that mass.  The accretion time is measured from the formation of an object (i.e., the insertion of a sink particle) to the {\it last} time at which its accretion rate drops below $10^{-7}$ M$_\odot$/yr, or the end of the calculation (which ever occurs first).  We also define an ejection time, which is the time between the formation of an object and {\it last} time the magnitude of its acceleration drops below 2000 km/s/Myr (or the end of the calculation).  The acceleration criterion is based on the fact that once an object is ejected from a stellar cluster through a dynamical encounter, its acceleration will drop to a low value.  The specific value of the acceleration was chosen by comparing animations and graphs of acceleration versus time for individual objects.  As with Calculations 1--3, the time-averaged accretion rates of the objects have a significant dispersion.  However, there is no systematic trend for the lower-mass objects to have lower time-averaged accretion rates.

\begin{figure}
    \includegraphics[width=8.8cm]{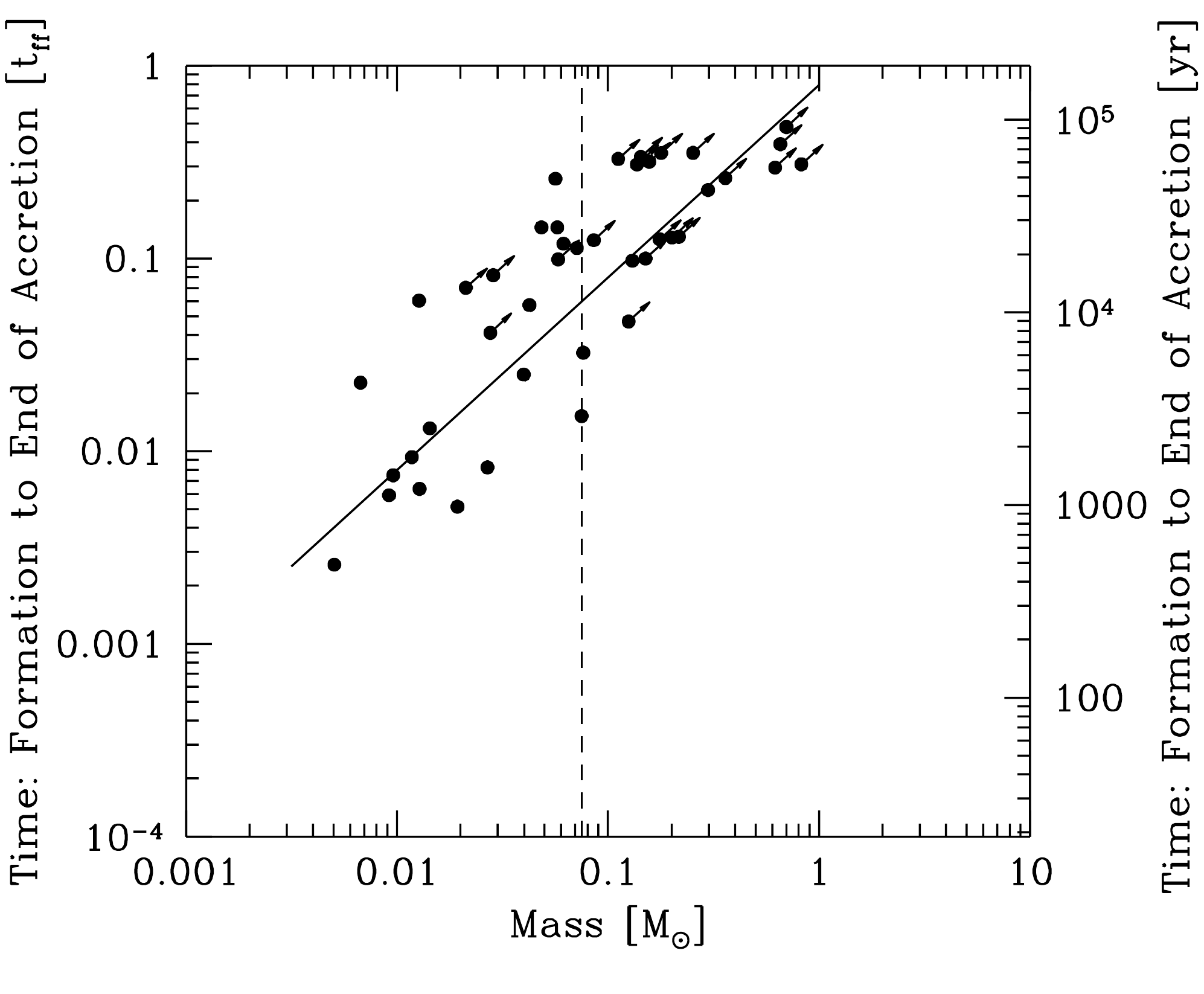}
\caption{\label{accmass} The time between the formation of each object and the termination
of its accretion or the end of the calculation versus its final mass.  As for the calculations in BB2005, there is 
a clear linear correlation between the time an object spends accreting and its final mass.  
The solid line gives the curve that the objects would lie on if each object accreted at the 
mean of the time-averaged accretion rates.
The accretion times are given in units of the $t_{\rm ff}$ on the left-hand axes
and years on the right-hand axes. The vertical dashed line marks the star/brown dwarf boundary.} 
\end{figure}

In Figure \ref{accmass}, we plot the time between the formation of an object and the termination of its accretion (or the end of the calculation) versus the final mass of the object.  Those points with arrows denote those objects that are still accreting significantly at the end of the calculation.  Accreting objects would move towards the upper right of the diagrams if the calculations were extended.  As with Calculations 1--3, it is clear that {\it the lower the final mass of the object, the earlier its accretion was terminated}.  This is the origin of the mass distribution of the objects: competition between accretion and ejection.

\begin{figure}
    \includegraphics[width=8.8cm]{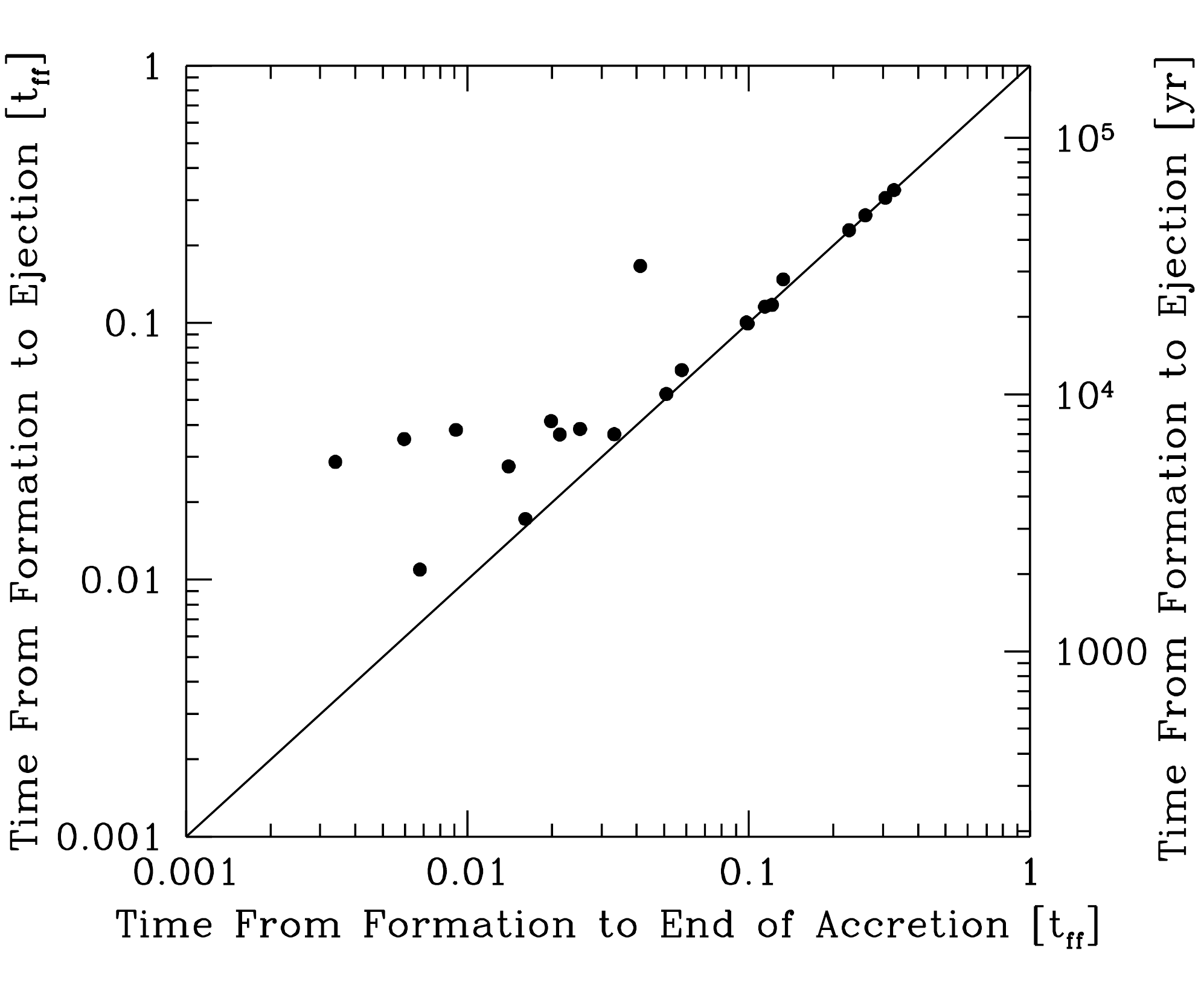}
\caption{\label{ejectacc} For each object that has stopped accreting, we plot the time between the formation of the object and its ejection from a multiple system versus the time between its formation 
and the termination of its accretion.  As for the calculations in BB2005, these times are correlated, showing that the termination of accretion on to an object is usually associated with dynamical ejection of the object. } 
\end{figure}

To check that the termination of the accretion is caused by the ejection of objects during dynamical interactions, in Figure \ref{ejectacc}, we plot the time between the formation of an object and its ejection from a stellar group versus the time between the formation of an object and the termination of its accretion.  In this figure, we only plot those objects that have stopped accreting and reached their final masses by the end of the calculations.  As in Calculations 1--3, the ejection and accretion times are closely correlated, showing that {\it the termination of accretion on to an object is usually associated with dynamical ejection of the object}.  These results confirm the speculation of \cite{ReiCla2001} and the conclusions of \cite{BatBonBro2002a} and BB2005 that brown dwarfs are `failed stars'.  They fall short of reaching stellar masses because they are cut off from their source of accretion prematurely due to ejection in dynamical interactions.

\begin{table*}
\begin{tabular}{lcccccl}\hline
Object Numbers & $M_1$ & $M_2$  & $q$ & $a$  & $e$  & Comments \\
        & M$_\odot$ & M$_\odot$ &  & &  \\ \hline
34,39    & 0.12  & 0.028  & 0.22  & ~~1.3*   & ~~0.92*  & Star-brown dwarf binary\\ 
8,10    & 0.18  & 0.14  & 0.80  & ~~2.2*   & ~~0.16*  & In Core 1\\ 
16,19   & 0.82  & 0.62  & 0.75  & ~~2.1*   & ~~0.89*  & \\ 
7,14   & 0.25  & 0.16 & 0.62  & ~~4.0*   & ~~0.48*  & \\
1,4   & 0.70  & 0.65 & 0.94  & ~~4.2*   & ~~0.73*  & \\
30,31   & 0.013  & 0.010 & 0.75  & 26   & 0.75  & Binary brown dwarf, ejected from Core 1 \\
24,28   & 0.17  & 0.057 & 0.33  & 10   & 0.46  & Star-brown dwarf binary \\
21,25   & 0.36  & 0.085 & 0.24  & 21   & 0.35  & \\
2,11   & 0.30  & 0.13 & 0.44  & 3560   & 0.97  & Wide binary, ejected from Core 1 \\ \hline
(7,14),12 & (0.41) & 0.11 & 0.27 & 38    & 0.93  &  \\
(16,19),27 & (1.44) & 0.15 & 0.10 & 40    & 0.32  & \\  \hline
(1,4),(34,39) &(1.35)&(0.15)& 0.11 & 69  & 0.38  &  In Core 1 \\
((7,14),12),17 & (0.52) & 0.14 & 0.26 & 182    & 0.54  & In Core 2  \\
(21,25),(24,28) &(0.44)&(0.23)& 0.52 & 154  & 0.23  &  \\
((16,19),27),38 & (1.59) & 0.012 & 0.007 & 1324    & 0.87  & Wide brown dwarf companion;  In Core 1 \\ \hline
((21,25),(24,28)),33 &(0.68)& 0.021& 0.03 & 281  & 0.90  & In Core 2  \\  \hline
\end{tabular}
\caption{\label{table3} The properties of the 7 multiple systems with semi-major axes less than 2000 AU and a very wide binary (3560 AU) formed in Calculation 4 (see also Figure \ref{binaryq}).  Three of these systems are binaries, three are quadruples, while the other is a quintuple system.  In addition, the three systems labelled `In Core 1' are mutually bound, as are the two systems `In Core 2'.  The structure of each system is described using a binary hierarchy.  For each `binary' we give the masses of the primary $M_1$ and secondary $M_2$, the mass ratio $q=M_2/M_1$, the semi-major axis $a$, and the eccentricity $e$.  The combined masses of multiple systems are given in parentheses.  Orbital quantities marked with asterisks are unreliable because these close binaries have periastron distances less than the gravitational softening length.  When the calculation is stopped, all but the ejected binary brown dwarf (30,31) are unstable and/or are still accreting, so their final states are unknown.}
\end{table*}

\begin{figure}
    \includegraphics[width=8.2cm]{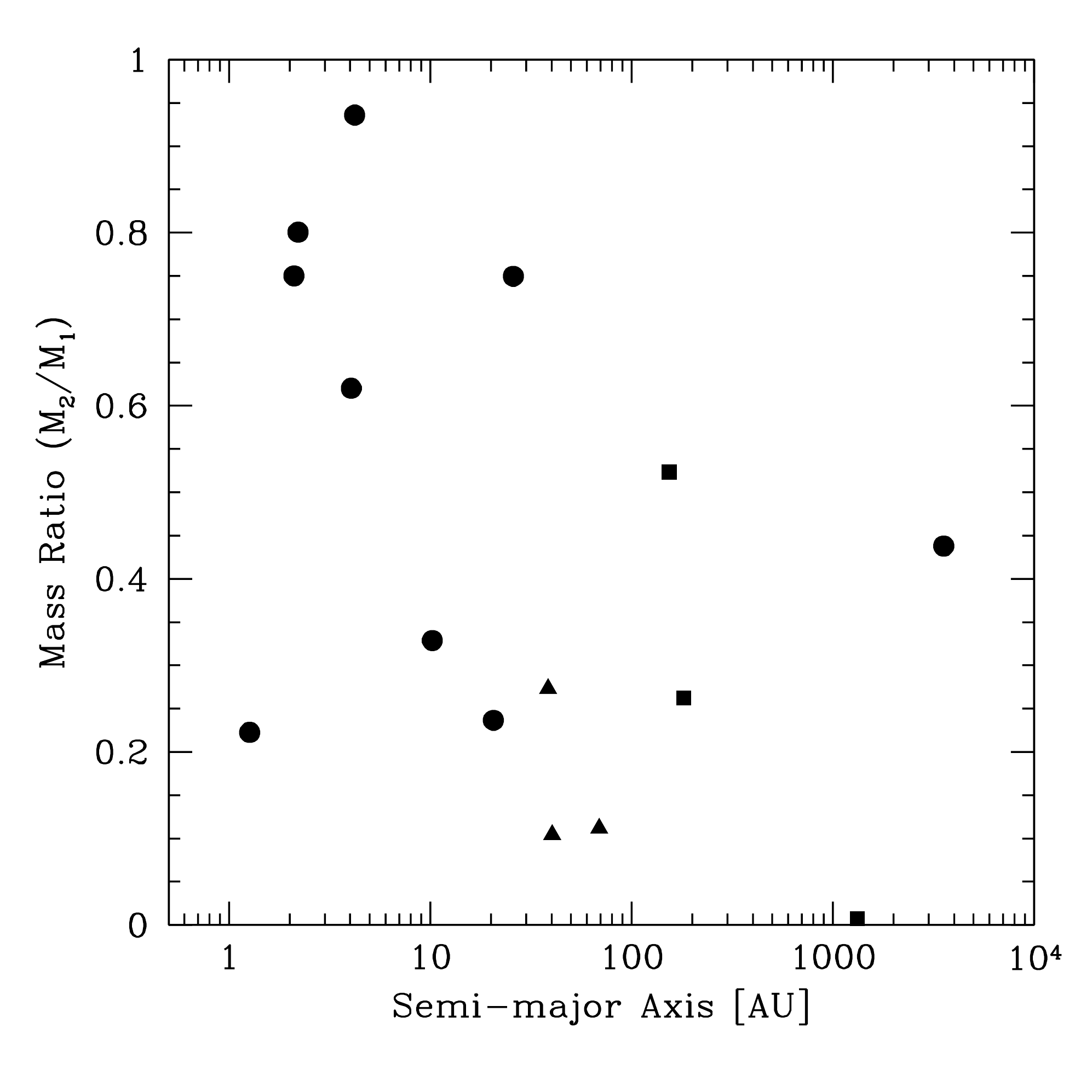}
\caption{\label{binaryq} Mass ratios versus semi-major axes of the binary, triple and
quadruple systems that exist at the end of the calculation (see also Table 3).  Binaries are plotted 
with circles, triples with triangles and quadruple systems with squares.  This figure should be
compared with Figure 12 of BBB2003, Figure 11 of BB2005, and Figure 12 of B2005 for the equivalent results from the other three calculations.  Calculation 1 produced no wide binaries (separations 
$>10$ AU) and no binaries with mass ratios $M_2/M_1 \la 0.3$. Calculation 2 produced
five wide binaries and three binaries with mass ratios $M_2/M_1<0.2$.  Calculation 3 produced two wide binaries and one binary with a mass ratio $M_2/M_1<0.3$. This calculation produces four wide binaries (separations $>10$ AU), and two binaries with mass ratios $M_2/M_1<0.3$.}
\end{figure}

\subsection{Multiple systems}
\label{multiplesystems}

In all four calculations, the dominant formation mechanism for binary and multiple systems 
is fragmentation, either of gaseous filaments \citep[e.g.][]{Bastien1983, Bastienetal1991, InuMiy1992} or of massive circumstellar discs (e.g., \citealt{Bonnell1994, BonBat1994a, Whitworthetal1995, BurBatBod1997, Hennebelleetal2004, BatBonBro2002a}).  Star-disc encounters play an important role in truncating discs (Section \ref{ppdiscs}), and in dissipating kinetic energy \citep[c.f.][]{Larson2002}, but they do not play a significant role in forming binary and multiple systems from unbound objects \citep[c.f.][]{ClaPri1991a}.  Only two star-disc encounters resulted in the formation of multiple systems in Calculation 1 and three in Calculation 3, while in Calculation 2 there was no obvious example of a multiple system being formed via a star-disc encounter.  In Calculation 4, there are two occurrences of star-disc encounters resulting in the formation of multiple systems when objects form along filaments in the first dense core and fall along the filament to form multiple systems.

\begin{figure*}
    \includegraphics[width=15.8cm]{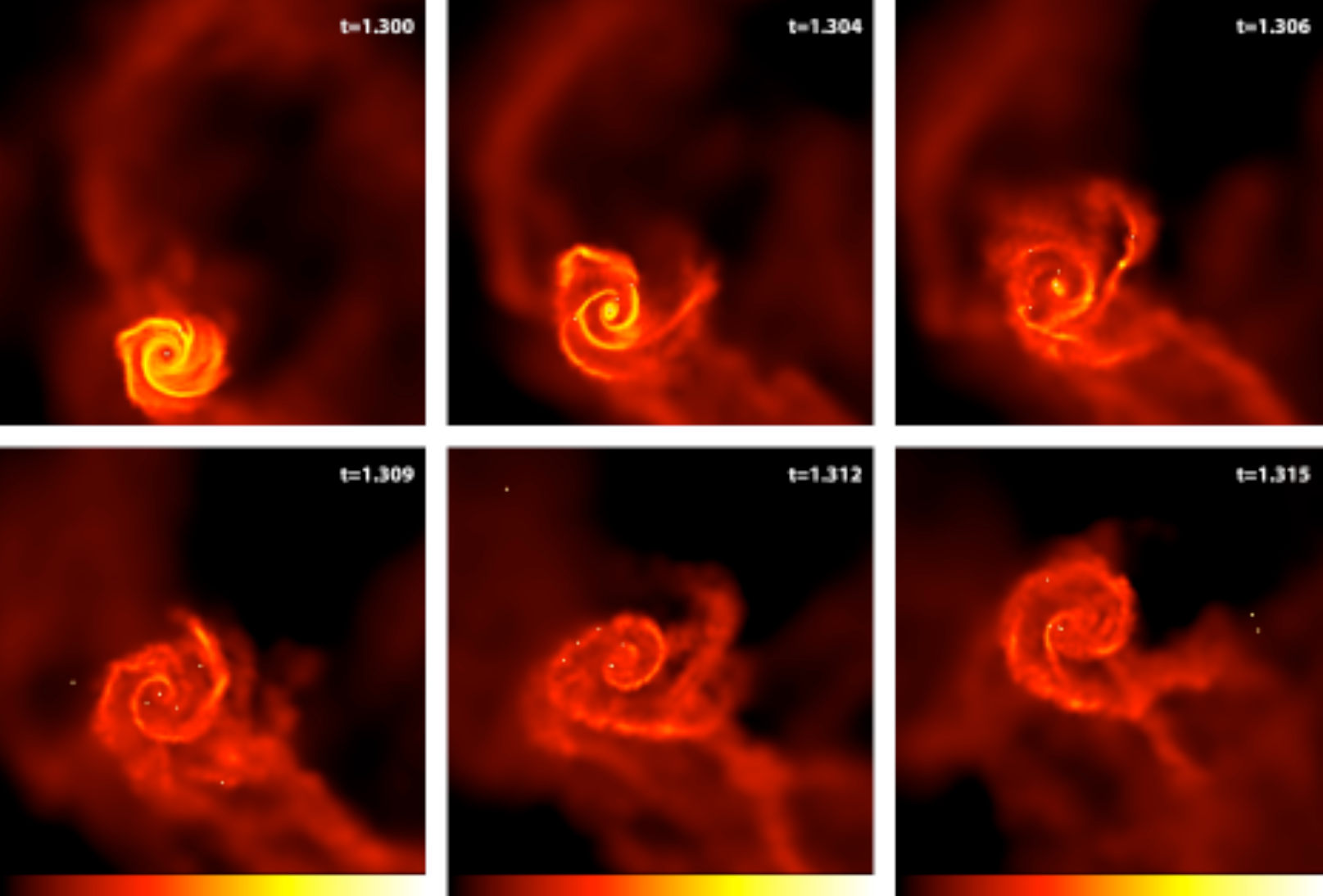}
\caption{\label{bbdfig} The formation of an ejected binary brown dwarf from the fragmentation of a massive circumstellar disc.  A massive ($\approx 0.6$~M$_\odot$ of gas) disc around a close binary system (composed of 0.42~M$_\odot$ and a 0.32~M$_\odot$ stars) fragments to form a low-mass star and four brown dwarfs ($t=1.300-1.307~t_{\rm ff}$). The two objects forming at the right-hand side of the disc in the $t=1.306~t_{\rm ff}$ panel end up as a brown dwarf binary with a semi-major axis of 26 AU that is ejected from the multiple system (right-hand side of the $t=1.315~t_{\rm ff}$ panel).  Each panel is 700 AU across.  Time is given in units of the initial free-fall time of $1.90\times 10^5$ years.  The panels show the logarithm of column density, $N$, through the cloud, with the scale covering $0.0 < \log N < 4.0$ with $N$ measured in g cm$^{-2}$. }
\end{figure*}

\subsubsection{Multiplicity}

When Calculation 4 was stopped, there were 19 single objects, 2 binaries, 3 quadruple systems, and 1 quintuple system (taking any objects with semi-major axes greater than $2000$ AU to be essentially unbound).  In addition, there was one very wide (3560 AU semi-major axis) highly eccentric ($e=0.97$) binary.  The properties of the 7 multiple systems are displayed in Table \ref{table3} and in Figure \ref{binaryq}.  Five of these systems originated in the first dense core.  Three of the multiple systems that originated in Core 1 are still weakly bound to each other, but the ejected binary brown dwarf system (30,31) and the very wide binary system (2,11) are unbound.  The two multiple systems in Core 2 are also marginally bound to each other.

Calculation 4 produces a high companion star fraction
\begin{equation}
CSF=\frac{B+2T+3Q+...}{S+B+T+Q+...}
\end{equation}
of $16/26 = 62$ percent, where $S$ is the number of single stars, $B$ is the number of binaries, $T$ is the number of triples, etc.  Alternately, the number of companions divided by the total number of objects is $16/42=38$ percent.  These percentages are similar to those of Calculations 1 and 3 and somewhat lower than Calculation 2.  Although the systems with more than two components will continue to evolve and some will probably eject more members, it is plausible that the final companion star frequency will be high, as required by observations of star-forming regions (\citealt*{GheNeuMat1993}; \citealt{Leinertetal1993,ReiZin1993}; \citealt{Richichietal1994}; \citealt{Simonetal1995}; \citealt{Ghezetal1997}; \citealt{Duchene1999}).

As with Calculations 1--3, Calculation 4 produces a realistic frequency of close binaries (separations $<10$ AU) even though no two objects form closer than 14 AU from each other due to the opacity limit for fragmentation (see \citealt{BatBonBro2002b} for a full discussion of how accretion, dynamical interactions with discs, and dynamical encounters can produce close binaries). Even if all wider systems break up, the resulting frequency of close binaries would be $5/37\approx14$ percent.  The corresponding values from Calculations 1--3 were 16 percent, 7 percent, and 10 percent, respectively.  The observed value is $\approx 20$ percent (Duquennoy \& Mayor 1991).  However, Duquennoy \& Mayor were not sensitive to brown dwarfs.  If only stars are considered, the frequency of close binaries becomes $4/16\approx 25$ percent (for Calculations 1--3 the corresponding frequencies were similar at $5/18\approx 28$ percent, $4/15\approx 27$ percent, and $3/13\approx 23$ percent, respectively).  As in Calculations 1--3, there is a preference for close binaries to have equal masses (all but the star-brown dwarf close binary have mass ratios of $M_2/M_1>0.6$), and the frequency of close binaries is higher for more massive primaries -- 9 of the 20 stars are members of close binaries, while only one brown dwarf is in a close binary.  These preferences result from the formation mechanisms of close systems as discussed by \cite{BatBonBro2002b}.

\subsubsection{Brown-dwarf companions to stars and brown dwarfs}

Together, Calculations 1--3 produced 6 binaries consisting only of very-low-mass (VLM) stars ($M<0.09$ M$_\odot$) or brown dwarfs out of $\approx 100$ VLM or brown dwarf systems, implying a frequency of binary brown dwarfs of $\sim 6$ percent.  Calculation 4 is consistent with this frequency in that it produced one VLM binary from $\approx 23$ VLM objects -- the system (30,31) which has a 26 AU semi-major axis and consists of 13 and 10 M$_{\rm J}$ brown dwarfs (Table \ref{table3}).  This binary brown dwarf system formed in the first dense core via the fragmentation of a gravitationally unstable disc surrounding a 4-AU binary star system consisting of a 0.42~M$_\odot$ and a 0.32~M$_\odot$ star (see Figure \ref{bbdfig}).  The disc fragmented into  five objects: a 0.15~M$_\odot$ low-mass star, a 75 M$_{\rm J}$ object right at the star/brown dwarf boundary, a 19 M$_{\rm J}$ brown dwarf, and the binary brown dwarf itself.  All these objects were dynamically ejected from the multiple system except the 0.15~M$_\odot$ low-mass star which, together with the close binary survives until the end of the calculation (see system (16,19),27 in Table \ref{table3}).  Note that both components of the close binary each almost double in mass between the disc fragmentation event and the end of the calculation; 0.3~M$_\odot$ of this extra mass comes is accreted from the massive disc during the disc fragmentation and dynamical rearrangement process.  The disc contained about 0.6~M$_\odot$ when it began to fragment (3/4 of the mass of the central close binary).

Disc fragmentation to form binary and multiple stars has been the topic of many numerical studies \citep{Bonnell1994, BonBat1994a, Whitworthetal1995, BurBatBod1997, Hennebelleetal2004}.  More recently, their role in the formation of brown dwarfs has been highlighted.  \citet{BatBonBro2002a} found that 3/4 of the brown dwarfs formed in their simulation of star cluster formation originated from the fragmentation of massive protostellar discs and although may were ejected from the resulting multiple systems, none of the ejected systems were binary brown dwarfs.  Calculation 2 (BB2005) produced three ejected binary brown dwarfs (all wide systems with semi-major axes $>60$ AU), but none of these were formed in a single disc fragmentation event.  Finally, \cite{StaWhi2009} performed 12 hydrodynamical simulations, including radiative cooling, of stars surrounded by massive protostellar discs that fragmented to produce many low-mass stars and brown dwarfs including two ejected binary brown dwarfs.  The formation of the binary brown dwarf depicted in Figure \ref{bbdfig} is similar to the formation of the two binary brown dwarfs produced in \citeauthor{StaWhi2009}'s calculations, except that the disc here is still accreting from the larger envelope as it fragments.

\begin{table*}
\begin{tabular}{lll}\hline
Disc Radius & Encircled Objects & Comments \\
\multicolumn{1}{c}{AU}        &         &          \\\hline
500         & ((21,25),(24,28)),33 & Circumquintuple disc (Figure \ref{core2}, $t=1.40 t_{\rm ff}$, central multiple system) \\ 
170         & (7,14),12 & Circumtriple disc (Figure \ref{core2}, $t=1.40 t_{\rm ff}$, upper multiple system) \\ 
160         & 23 & Circumstellar disc around 0.20 ${\rm M}_\odot$ star (Figure \ref{core1}, $t=1.40 t_{\rm ff}$, disc at lower left of main core)\\ 
150         & (8,10) & Circumbinary disc,  (Figure \ref{core2}, $t=1.40 t_{\rm ff}$, disc at bottom) \\
100         & (1,4),(34,39)    & Very disturbed circumquadruple disc (Figure \ref{core1}, $t=1.40 t_{\rm ff}$, lower system in the main core) \\ 
90           & 22 & Circumstellar disc around 0.22 ${\rm M}_\odot$ star  (Figure \ref{core1}, $t=1.40 t_{\rm ff}$, disc at lower right of main core) \\ 
80         & (16,19),27    & Very disturbed circumtriple disc (Figure \ref{core1}, $t=1.40 t_{\rm ff}$, upper system in the main core) \\ 
 \hline
\end{tabular}
\caption{\label{tablediscs} The discs that exist around objects when the calculation is stopped. Discs with radii $\lsim 10$ AU are not resolved.  Unlike Calculation 1, in Calculation 4 no objects are ejected with resolved discs. This table should be
compared with Tables 4 of BBB2003, BB2005, and B2005 for the equivalent results
from Calculations 1--3, respectively.}
\end{table*}

\begin{figure}
    \includegraphics[width=8.4cm]{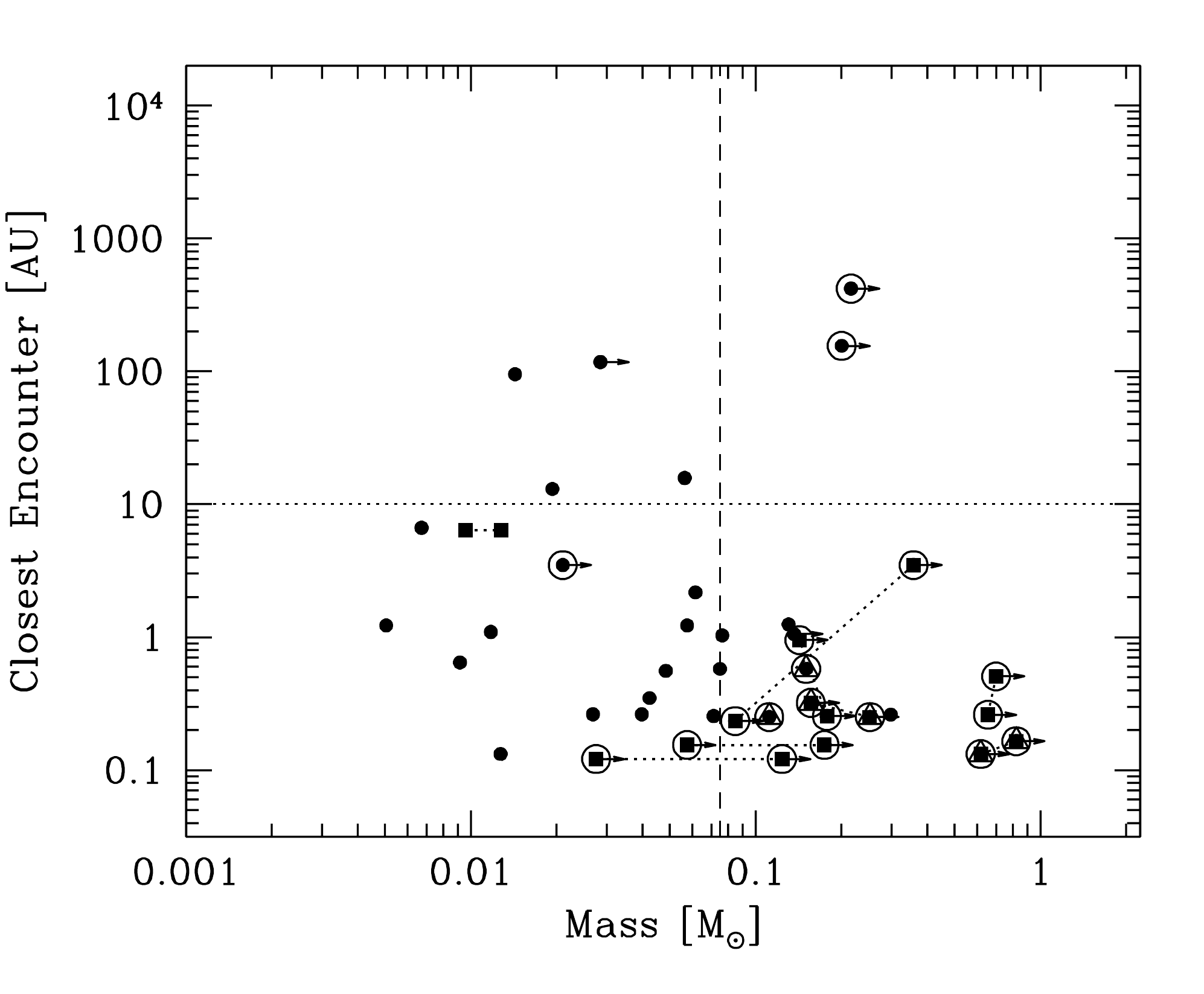}
\caption{\label{truncate}  The closest encounter distance of each 
star or brown dwarf during the calculation versus the object's final 
mass.  This figure should be compared with Figure 14 of BBB2003, 
Figure 12 of BB2005, and Figure 13 of B2005, for the equivalent results from 
Calculations 1, 2, and 3.  Objects that 
are still accreting significantly at the end of the calculation
are denoted with arrows indicating that they are still evolving and
that their masses are lower limits.  Objects that
have resolved discs at the end of the simulation are circled.
Discs smaller than $\approx 10$ AU (horizontal dotted line) cannot 
be resolved by the simulation.  Objects that have had close 
encounters may still have resolved discs due to subsequent accretion
from the cloud.  Note that there are only 7 resolved discs 
at the end of the simulation, but many surround binary and 
higher-order multiple systems (hence the 19 circles in the figure).
Binaries (semi-major axes 
$<100$ AU) are plotted with the two components connected by dotted 
lines and squares are used as opposed to circles.  
Components of triple systems whose orbits have semi-major 
axes $10<a<100$ AU are denoted by triangles.  All but one of the 
binaries is surrounded by a resolved disc (the binary brown dwarf is without a resolved disc).  
Encounter distances less than 4 AU are upper limits since the point 
mass potential is softened within this radius.  The vertical 
dashed line marks the star/brown dwarf boundary.  There are no brown 
dwarfs that have resolved discs and have finished accreting.  The brown 
dwarf at the upper left of the graph was formed during the fragmentation 
of a circumstellar disc into several low-mass objects and ejected 
without a resolved disc despite having a closest encounter distance of 95 AU.} 
\end{figure}

The observed frequency of very-low-mass and brown dwarf binaries is $\approx 20$ percent \citep{Reidetal2001, Closeetal2002, Closeetal2003, Bouyetal2003, Burgasseretal2003, Gizisetal2003, Martinetal2003, Siegleretal2005, Burgasseretal2006, BasRei2006, Reidetal2006, Closeetal2007, Ahmicetal2007, Reidetal2008}.  Thus, at face value, Calculations 1--4 tend to under-produce binary brown dwarfs. However, \cite{Bate2009a} recently performed a calculation similar to Calculation 1, but of a cloud an order of magnitude more massive that produced more than 800 brown dwarfs.  Although overall this calculation produced a similar very-low-mass binary frequency to the smaller calculations, with the better statistics it was possible to sub-divide the population.  This showed that the binary frequency decreases strongly and monotonically with decreasing primary mass and that for very-low-mass stars and high-mass brown dwarfs (those typically targeted by observational surveys) the binary frequency was in better agreement with observations.  Even better agreement with observations was obtained when \citet{Bate2009a} repeated the calculation with smaller sink particle accretion radii (0.5 AU rather than 5 AU) and without gravitational softening.  Thus, the apparent disagreement with observations is due to both small number statistics (which make it necessary to calculate the binary frequency over a wide range of primary masses) and the fact that the calculation does not resolve circumstellar discs at radii $\la 10$ AU and softens gravitational interactions between stars/brown dwarfs at separations less than 4 AU.  It is not a fundamental failing of hydrodynamical star formation simulations.

For star-brown dwarf binary systems, the frequencies are also very low.  Calculation 1 one produced one binary system consisting of a star (0.13 M$_\odot$) and a brown dwarf (0.04 M$_\odot$).  The system had a separation of 7 AU and was part of an unstable septuple system.  Both objects were still accreting.  Calculations 2 and 3 did not produce any such star/brown dwarf binary systems.  Calculation 4 produces two systems (see 34,39 and 24,28 in Table \ref{table3}).  Both are members of higher-order systems and are still accreting so may not survive as star-brown dwarf binaries.  The reasons for the low frequency of star-brown dwarf binaries are discussed by BB2005.   The rarity of brown dwarfs orbiting stars is consistent both with the so called brown dwarf desert discovered through Doppler searches for planets orbiting solar-type stars \citep{MarBut2000, GreLin2006} and from imaging surveys for wide systems \citep{Gizisetal2001,McCZuc2004,MetHil2004,MetHil2009}.

\subsection{Protoplanetary discs}
\label{ppdiscs}

The calculations resolve gaseous discs with radii $\ga 10$ AU around the young stars and brown dwarfs.  Discs with typical radii of $\sim 100$ AU form around many of the objects due to the infall of gas with high specific angular momentum.  However, in all calculations many of the discs are severely truncated in subsequent dynamical interactions, leaving most of them too small to form analogues of our solar system (see BBB2003).  

The seven resolved discs at the end of Calculation 4 are listed in Table \ref{tablediscs}, and in Figure \ref{truncate} we plot the closest encounter distance for each object during the calculation as a function of its final mass and which of these objects is surrounded by a resolved disc.  Many of the discs are actually circumbinary or circum- multiple discs (see Table \ref{tablediscs}) and, in fact, all of the binaries are surrounded by resolved discs with the exception of the ejected binary brown dwarf (Figure \ref{bbdfig}) and the ejected wide binary (Tables \ref{table3} and \ref{tablediscs}). All but two stars have had encounters closer than 10 AU.  Although they have had very close encounters, subsequent infalling gas has build up new circumbinary and circum-multiple discs around most of them.  This is a feature of all four calculations.   Only three of the brown dwarfs have resolved discs, but these are brown dwarfs that are components of multiple systems surrounded by circum-multiple discs (objects 28, 33, 39 in Table \ref{tablediscs}) and they are also still accreting. All of the stars with resolved discs are members of multiple systems, except objects 22 and 23 (Table \ref{tablediscs}) in the upper right of Figure \ref{truncate}.  These are both bound to the small group of stars in the first dense core at the end of the calculation and are visible to the lower right and lower left, respectively, of the main core in the $t=1.40~t_{\rm ff}$ panel of Figure \ref{core1}.

\section{Discussion}

\subsection{The accretion/ejection model for the IMF}

BB2005 and B2005 argued that the IMFs produced by Calculations 1--3 originated from a combination of accretion and dynamical ejections which terminate the accretion.  Calculation 4 again supports this model in that Figure \ref{accrates} shows there is no correlation between an object's time-averaged accretion rate and its final mass, while Figure \ref{accmass} shows a strong correlation between the time an object spends accreting and its final mass and Figure \ref{ejectacc} shows that accretion is usually terminated by gravitational interactions with other objects leading to dynamical ejection.

The simple accretion/ejection model proposed by BB2005 for the IMF produced by a star-forming molecular cloud is as follows.  
\begin{itemize}
\item We assume all objects begin with masses set by the opacity limit for fragmentation ($3$ M$_{\rm J}$ for Calculations 1,2, and 4 and $9$ M$_{\rm J}$ for Calculation 3) and then accrete at a fixed rate $\dot{M}$ until they are ejected.
\item We assume the accretion rates of individual objects are drawn from a log-normal distribution with a mean accretion rate (in log-space) given by $\log_{10}(\overline{\dot{M}})=\overline{\log_{10}(\dot{M})}$ and a dispersion of $\sigma$ dex (i.e. $\log_{\rm 10}(\dot{M}) = \log_{\rm 10}(\overline{\dot{M}}) + \sigma G$, where $G$ is a random Gaussian deviate with zero mean and unity variance).
\item The ejection of protostars from an $N$-body system is a stochastic process that can be described in terms of the half-life of the process.  We assume that there is a single parameter, $\tau_{\rm eject}$, that is the characteristic timescale between the formation of an object and its ejection from the cloud.  The probability of an individual object being ejected is then $\exp(-t/\tau_{\rm eject})$ where $t$ is the time elapsed since its formation.
\end{itemize}

Assuming that the cloud forms a large number of objects, $N$, and that the time it evolves for is much greater than the characteristic ejection time, $T\gg \tau_{\rm eject}$, then a semi-analytic formula can be derived for the form of the IMF (BB2005) and there are essentially only three free parameters in the model.  These are the mean accretion rate times the ejection timescale, $\overline{M}=\overline{\dot{M}}\tau_{\rm eject}$, the dispersion in the time-averaged accretion rates, $\sigma$, and the minimum mass provided by the opacity limit for fragmentation, $M_{\rm min}$.  If $\overline{M}>>M_{\rm min}$, $\overline{M}$ is the characteristic mass of an object.  

However, the hydrodynamical calculations are not followed until all the stars and brown dwarfs have finished accreting (i.e., the IMF is not fully formed), so it is not the case that $T\gg \tau_{\rm eject}$.  This must be taken into account when calculating simple accretion/ejection models for comparison with the IMFs from the hydrodynamical calculations.  To do this, we evolved the simple models over the same periods of time, $T$, that the hydrodynamical simulations took to form their stars and brown dwarfs and take the times of formation of each of the objects directly from the hydrodynamical simulations (i.e., from Figure \ref{sfrate} for Calculation 4).

\begin{figure}
    \includegraphics[width=8.7cm]{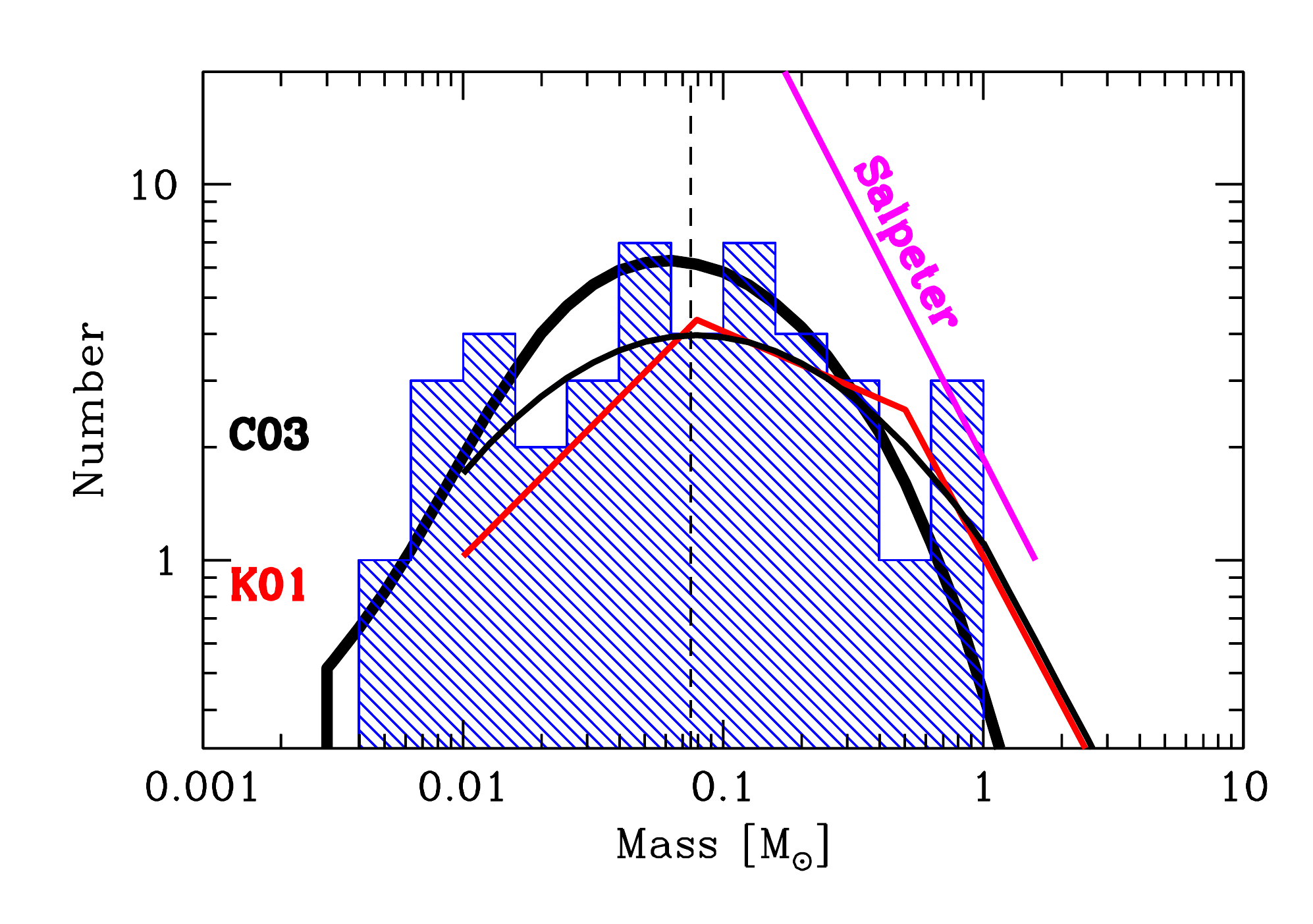}
\caption{\label{imffit} The initial mass functions produced by Calculation 4 (histogram) and its fit using the simple accretion/ejection IMF model (thick curve).  Statistically, the hydrodynamical and the model IMFs are indistinguishable (a Kolmogorov-Smirnov test gives a 42 percent probability that the hydrodynamical IMF could have been drawn from the model IMF).  Also shown are the \citet{Salpeter1955} slope (solid straight line), and the \citet{Kroupa2001} (solid broken line) and \citet{Chabrier2003} (thin curve) mass functions.  The vertical dashed line is the stellar-substellar boundary.} 
\end{figure}

\begin{table}
\begin{tabular}{lccccc}\hline
Model\hspace{-5pt} & $M_{\rm min}$ & $\overline{\dot{M}}$ & $\sigma$ & $\tau_{\rm eject}$  & $T$\\
       &  M$_\odot$ &  M$_\odot$/yr     &  Dex.    & yr  &  yr    \\\hline
1 & 0.003 & $6.17\times 10^{-6}$ & 0.33 & $3.2\times 10^4$ & $6.91\times 10^4$ \\
2 & 0.003 & $7.18\times 10^{-6}$ & 0.50 &  $9.3\times 10^3$ &   $3.67\times 10^4$\\
3 & 0.009 & $1.00\times 10^{-5}$ & 0.41 &  $2.5\times 10^4$ &   $6.91\times 10^4$\\
4 & 0.003 & $4.86\times 10^{-6}$ & 0.35 &  $3.7\times 10^4$ &   $9.11\times 10^4$\\
 \hline
\end{tabular}
\caption{\label{tableimf} The parameters of the simple accretion/ejection IMF models that should reproduce the IMFs from the four hydrodynamical calculations (Figure \ref{imffit}).  There are essentially three parameters in the models, the mean accretion rate times the characteristic timescale for ejection $(\overline{\dot{M}}\tau_{\rm eject})$, the dispersion in the accretion rates $\sigma$, and the minimum mass set by the opacity limit for fragmentation $M_{\rm min}$.  The time period over which the simulations are run, $T$, has a small effect on the form of the IMF.}
\end{table}

We then generated a model IMF for comparison with the IMF from Calculation 4.  The model IMF is the average of 30000 random realisations of the simple accretion/ejection model, keeping the values of the input parameters fixed.  The parameter values are given in Table \ref{tableimf}.  
It is important to note that these parameters were {\it not} varied in order to obtain good fits to the hydrodynamical IMF.  Rather, {\it the values of the parameters were taken directly from the hydrodynamical simulation}.  There is no freedom to vary the parameters in order to obtain a better fit. 
The mean accretion rate of the objects, $\overline{\dot{M}}$, and the dispersion in the accretion rates, $\sigma$, were set equal to the mean (in log-space) of the time-averaged accretion rates and their dispersion from Figure \ref{accrates}.  The characteristic ejection times, $\tau_{\rm eject}$, were set so that the mean numbers of objects ejected from the 30000 random realisations matched the number of objects ejected during each of the hydrodynamical calculations (24 for Calculation 4).

Figure \ref{imffit} shows that the simple accretion/ejection model matches the hydrodynamical IMF well.  A Kolmogorov-Smirnov test gives a 42 percent probability that the hydrodynamical IMF could have been drawn from the model IMF (i.e., they are statistically indistinguishable).  For Calculations 1--3, the hydrodynamical IMFs have probabilities of 92, 27, and 7 percent, respectively (see BB2005 and B2005).

This model for the origin of the IMF depends on dynamical interactions between protostars that primarily form in groups or clusters.  While such dynamical interactions may be common in high-density star-forming regions, recent observational studies of lower density star-forming regions such as Ophiuchus \citep{Andreetal2007} find that the velocity dispersion between cores is only a couple of times the sound speed implying that these cores do not have time to interact with each other before evolving into pre-main sequence objects.  However, this does not preclude a role for dynamical interactions in the shaping of the IMF if each core produces several protostars since competitive accretion and dynamical interactions and ejections between objects operate equally well in small groups \cite[e.g., the original competitive accretion paper of][]{Bonnelletal1997}.  Indeed, even in the very low-density Taurus star-forming region the pre-main sequence stars are observed to be in small ($\approx 1$ pc diameter) groups of $\sim 10$ objects \citep{Gomezetal1993}.  If each of these formed as more compact stellar groups (a distinct possibility given the current ages, sizes, and internal velocity dispersions of the groups), competitive accretion and dynamical interactions may have played an important role even in this very distributed star-forming region.

\subsection{The dependence of the IMF on the structure of molecular clouds}

Since the four hydrodynamical calculations discussed in this paper are time consuming, they have been carefully designed to enable the origins of the statistical properties of stars to be investigated in the most possible detail.  Comparison of Calculations 1 and 2 allowed BB2005 to investigate the dependence of star formation on the mean density $\bar{\rho}$ of the molecular cloud and, therefore, the mean thermal Jeans mass which scales as $1/\sqrt{\bar{\rho}}$.  Comparison of Calculations 1 and 3 allowed B2005 to investigate the roles of the opacity limit for fragmentation and the mean temperature, $T$, of a molecular cloud on the star formation process.  In agreement with BB2005, B2005 found that the characteristic stellar mass obtained from such calculations depends primarily on the mean thermal Jeans mass of a star-forming molecular cloud (which also scales as $T^{3/2}$) and not on the opacity limit for fragmentation.  The opacity limit for fragmentation was found only to set the low-mass cut-off of the IMF.

Calculation 4, discussed in this paper, is identical to Calculation 1 (BBB2003) except for the velocity field that is initially imposed on the cloud.  Two questions can be investigated by comparing the results of these two calculations.  The first is, given that only one random realisation of the velocity field was tried in Calculation 1, do we know that the results obtained were not in some sense `special'?  In other words, if Calculation 1 was performed, say, 10 times, how much would the results vary?  Although we cannot measure the dispersion in the results from only two calculations, if the results are statistically indistinguishable this gives us confidence that the results of Calculation 1 were not unusual somehow.  The second question we can investigate is whether the results are sensitive to the power spectrum of the imposed velocity perturbations.  The original power spectrum of $P(k)\propto k^{-4}$ used for Calculation 1 was motivated by the Larson scaling relations  \citep{Larson1981} for the observed velocity-size relation of turbulent motions in molecular clouds (see Section \ref{initialcond}).  Does using a very different initial power spectrum influence the results noticeably?

By comparing the results of Calculations 1 and 4, the answers to these questions is that there is no evidence that changing the imposed initial velocity field (either the particular random realisation or the power spectrum) has any effect on the statistics of the stellar systems produced during the star formation process.  The IMFs obtained from the two calculations are statistically indistinguishable.  Neither is there any statistically significant change in the other stellar properties such as velocity dispersion, multiplicity, or discs.  It is necessary for structure to be generated in the spherical cloud by the imposed velocity field \citep{ClaBon2005} so that the cloud does not collapse spherically-symmetrically to form a single massive object.  But beyond this, the specific cloud structure is not crucial to the results.  Rather, as long as small groups of stars and brown dwarfs are produced, competitive accretion (\citealt{Bonnelletal1997}, 2001a, 2001b) and ejection (\citealt{BatBonBro2002a}; BBB2003) act to determine the stellar properties (BB2005) independent of the cloud structure.

This is an important result for several reasons.  First, it is consistent with the observed insensitively of the star formation process to environment (see Section \ref{introduction}).  Observationally, the present-day IMF is surprisingly consistent from region to region within our Galaxy \citep{Scalo1998,Kroupa2002,Chabrier2003,ElmKleWil2008} and is often referred to as perhaps being `universal'.   If the IMF depended sensitively on cloud structure then one might expect to observe a variation of the IMF between different regions, perhaps depending on whether star formation was spontaneous within a molecular cloud or triggered by expanding HII regions or supernova shells.  However, if stellar properties are determined by competitive accretion and chaotic dynamical interactions between young stars on small scales in which memory of the initial conditions is quickly lost, as is the case here, this would help to explain the universality of the IMF.  Second, if the results of the star formation process are independent of the initial conditions, this may drastically reduce the difficulty of understanding the star formation process because we would not be limited by our understanding of the formation and evolution of molecular clouds, which is still very poor.  Finally, even if the initial conditions for calculations such as those discussed in this paper are very idealised and not particularly realistic, the lack of dependence of the stellar properties on the initial conditions increases our confidence that the results are meaningful. 

As mentioned in Section \ref{introduction}, \citet{DelClaBat2004} and \citet{GooWhiWar2006} also investigated the dependence of star formation on the initial power spectrum of the velocity field in hydrodynamical simulations.  They found contradictory results.  \citeauthor{DelClaBat2004} found that more low-mass objects were produced when more small-scale structure was present in the velocity field, while \citeauthor{GooWhiWar2006} found more low-mass objects in their calculations that began with more power on large-scales.  As discussed above, from Calculations 1 and 4 considered in this paper we find no evidence for a dependence of the IMF on the the initial imposed kinetic power spectrum.  Although there were differences between the two earlier studies (the latter began with less turbulence overall than the former) and there are also differences between the earlier studies and that presented here (the earlier studies were of isolated dense molecular cloud cores whereas the simulation here is of a larger-scale cloud), we conclude that the most likely interpretation is simply that the tendencies observed from both previous studies were not statistically significant.  \citeauthor{DelClaBat2004} explicitly state that their result was at the $2-\sigma$ level of significance.

In summary, when combined, the four hydrodynamical calculations from BBB2003, BB2005, B2005 and this paper show that the stellar properties obtained from hydrodynamical star formation calculations of the collapse of molecular clouds do not depend strongly on the onset of the opacity limit for fragmentation or the initial `turbulent' velocity field and the resulting cloud structure.  Rather, the characteristic mass of the IMF is set by the mean thermal Jeans mass in the progenitor cloud (determined by the cloud's density and temperature).  All these calculations have assumed decaying turbulence with an initial kinetic energy equal in magnitude to the gravitational energy of the cloud.  Other studies have shown that the star formation process and the IMF may also depend on the level of turbulence (\citealt{ClaBon2004,GooWhiWar2004a,Clarketal2005,Ballesterosetal2006,ClaBonKle2008}) in the cloud and whether or not the turbulence is driven and on what scales it is driven (\citealt{Klessen2001,Klessenetal2005, OffKleMcK2008}; \citealt{Offneretal2008}).  The equation of state of molecular gas may also play an important role in setting the Jeans mass in the clouds and, therefore, the characteristic stellar mass (\citealt{Larson2005,Jappsenetal2005}; \citealt*{BonClaBat2006}; \citealt{ElmKleWil2008}).  Recently, the effects on the IMF of radiative feedback \citep{KruKleMcK2007,Bate2009b}, magnetic fields \citep{PriBat2008}, and the two in combination \citep{PriBat2009} have begun to be investigated.  These studies show that radiative feedback and magnetic fields both tend to increase the characteristic stellar mass.

\section{Conclusions}

We have presented results from a hydrodynamical calculation to follow the collapse of a molecular cloud with decaying turbulence to form a stellar cluster while resolving fragmentation down to the opacity limit.  The calculation differs from the similar calculation performed by \citet{BatBonBro2003} in that the initial turbulent velocity field imposed on the cloud has a steeper power spectrum with more power being injected on large scales.  We compare the results with those obtained from the calculations published by \citet{BatBonBro2002a, BatBonBro2002b,BatBonBro2003}, \citet{BatBon2005}, and \citet{Bate2005}.

We find that although the power spectrum of the initial velocity field is very different ($P(k)\propto k^{-6}$ rather than $P(k)\propto k^{-4}$), the statistical properties of the stars and brown dwarfs formed during the calculation are statistically indistinguishable from the results reported by \citet{BatBonBro2003}.  In particular, a Kolmogorov-Smirnov test performed on the two IMFs shows that they are consistent with being drawn from the same underlying population.  We attribute the independence of the resulting stellar properties to the fact that the stars and brown dwarfs are formed in small groups which evolve primarily due to competitive accretion and chaotic dynamical interactions and ejections.  The fact that the velocity fields and density structure of the gas in the molecular clouds may differ does not significantly affect the result of these processes.

This result clearly demonstrates that the statistical properties of the stars and brown dwarfs produced by such hydrodynamical calculations of star cluster formation are relatively insensitive to the initial conditions.  This is a positive outcome in the sense that although the initial conditions for all such calculations are highly idealised, it gives us confidence that the evolution may still be representative of the star formation process.  The independence of the results of the star formation process to the kinematic structure of the molecular gas is also consistent with the fact that the IMF and other stellar properties are observed to be relatively `universal' properties that do not seem to vary significantly between different star-forming regions or depend on environment.  For example, if the products of the star formation process did depend sensitively on structure in the molecular gas then one might expect stellar properties to differ between regions of spontaneous star formation and star formation triggered in the gas swept up by supernova shells or the expansion of HII regions.  

Together, the four hydrodynamical calculations presented in this series of papers argues for the main determinant of stellar properties being the mean thermal Jeans mass in the molecular cloud upon which the characteristic mass of the IMF depends linearly \citep{BatBon2005}.  Changes in the opacity limit for fragmentation \citep{Bate2005} and the initial power spectrum of the velocity field of the gas (this paper) do not significant affect the statistical properties of the stars and brown dwarfs produced.

\section*{Acknowledgments}

MRB thanks the anonymous referee for suggesting clarifications to the original manuscript.  The computations reported here were performed using the U.K.
Astrophysical Fluids Facility (UKAFF).  Analysis of the calculation was performed using the University of Exeter supercomputer, an SGI Altix ICE 8200.  MRB is grateful for the support of a Philip Leverhulme Prize and a EURYI Award.  This work, conducted as part of the award ÒThe formation of stars and planets: Radiation hydrodynamical and magnetohydrodynamical simulationsÓ made under the European Heads of Research Councils and European Science Foundation EURYI (European Young Investigator) Awards scheme, was supported by funds from the Participating Organisations of EURYI and the EC Sixth Framework Programme.

\bibliography{mbate}

\end{document}